\documentclass[preprint]{aastex}
\usepackage{amsmath}
\usepackage{xspace}

\def\ie{{i.e.}}

\def\ltsima{$\; \buildrel < \over \sim \;$}
\def\simlt{\lower.5ex\hbox{\ltsima}}
\def\gtsima{$\; \buildrel > \over \sim \;$}
\def\simgt{\lower.5ex\hbox{\gtsima}}
\def\fesc{{$\langle f_{esc}\rangle$}\xspace}

\received{.......}
\accepted{.......}
\slugcomment{to be submitted to ApJ }

\begin{document}
\tighten
\thispagestyle{empty}
\pagestyle{myheadings}
\markright{DRAFT: \today\hfill}

\def\placefig#1{#1}

\title{FEEDBACK FROM GALAXY FORMATION:   
ESCAPING IONIZING RADIATION FROM GALAXIES AT HIGH REDSHIFT } 

\author{MASSIMO RICOTTI AND J. MICHAEL SHULL$^1$ } 
% .................................................................
\affil{
Center for Astrophysics and Space Astronomy \\
Department of Astrophysical and Planetary Sciences \\
University of Colorado, Campus Box 389, Boulder CO 80309 \\
E--mail: ricotti@casa.colorado.edu,mshull@casa.colorado.edu \\ 
$^1$ also at JILA, University of Colorado and National
Institute of Standards and Technology }  
%  .................................................................
 
\begin{abstract}
  Several observational and theoretical arguments suggest that
  starburst galaxies may rival quasars as sources of metagalactic
  ionizing radiation at redshifts $z > 3$.  Reionization of the
  intergalactic medium (IGM) at $z > 5$ may arise, in part, from the
  first luminous massive stars. To be an important source of radiative
  feedback from star formation, a substantial fraction ($\sim 1-10$\%)
  of the ionizing photons must escape the gas layers of the galaxies.
  Using models of smoothly distributed gas confined by dark-matter
  (DM) potentials, we estimate the fraction, \fesc, of Lyc flux that
  escapes the halos of spherical galaxies as a function of their mass
  and virialization redshift.  The gas density profile is found by
  solving the equation of hydrostatic equilibrium for the baryonic
  matter in the potential well of a DM halo with the density profile
  found by \cite{Navarro:96}. We then perform a parametric study to
  understand the dependence of \fesc on redshift, mass, baryonic
  fraction, star-formation efficiency (SFE), stellar density
  distribution, and OB association luminosity function. We give useful
  analytical formulas for $\langle f_{esc}(z,M_{DM})\rangle$.  Using
  the Press-Schechter formalism, we find that stellar reionization at
  $z \sim 7$ is probably dominated by small galactic sub-units, with
  $M_{DM} \simlt 10^7$ M$_\odot$ and SFE $\sim 10$ times that in the
  Milky Way. This may affect the distribution of heavy elements
  throughout the intergalactic medium.
\end{abstract}

\section{Introduction}

In all cosmological models, the temperature of the cosmic background
radiation at redshift $z \sim 1000$ is sufficiently low that hydrogen
ions recombine and radiation decouples from matter. The baryonic Jeans
mass after this event is $\sim 10^5$~M$_\odot$, and the first luminous
objects in a CDM cosmology could form at redshift $z\simeq 30$
\citep{Peebles:68,Binney:77,Rees:77,Silk:77,White:78,Tegmark:96,
  AbelBryan:98}. If fragmentation occurs, triggered by H$_2$ formation
and radiative cooling \citep{Lepp:84, AbelAnninos:98}, massive stars
are likely to form.  Depending on the slope of the initial mass
function (IMF) and on the star-formation efficiency (SFE), the
massive-star population will be a source of mechanical energy, heavy
elements, and Lyman continuum (Lyc) photons.  Such processes are known
as ``feedback'' from galaxy formation, and they can have substantial
impact on the intergalactic medium (IGM) and on future generations of
stars and galaxies.

In this paper, we focus on radiative feedback in the form of ionizing
radiation from massive stars.  In addition to its effects on diffuse
gas in the halos of galaxies, Lyc radiation from the first stars can
also provide an important source for ionizing the surrounding IGM
\citep{Madau:96}.  Observations of the transmitted flux below the
Ly$\alpha$ emission line in high-redshift quasars (the Gunn-Peterson
effect) imply that the diffuse IGM was already ionized at redshift $z
= 5$.  Although QSOs play a dominant role in photoionizing the IGM at
$z < 4$ \citep{Shapiro:87, Donahue:87, Meiksin:93}, their dwindling
numbers at $z > 4$ \citep{Pei:95, Miralda:90, Madau:98, Fardal:98}
suggest the need for another ionization source.  Unless a hidden
population of quasars is found, radiation emitted by high-redshift
massive stars seems necessary to reionize the universe. This
hypothesis is reinforced by recent observations of He~II absorption
\citep{Reimers:97}, from the redshift evolution of the column density
ratio of Si~IV/C~IV and C~II/C~IV in the Ly$\alpha$ forest
\citep{Songaila:96,Boksenberg:97}, and form the thermal history of
the IGM \citep{Ricotti:00a, Schaye:00} that favor a soft spectrum of the
ionizing radiation, more typical of hot stars than of quasars
\citep{Haardt:96, Fardal:98}.

A key ingredient in determining the effectiveness by which starburst
galaxies photoionize the surrounding IGM is the parameter \fesc,
the escaping fraction of Lyc photons.  This factor depends on the
shape of the galaxy, on the gas and stellar density profiles, 
and on the IMF and SFE.  Recent estimates \citep{Madau:96} 
suggest that if 10\% of the
ionizing photons from hot stars escape the disk gas layers [\fesc
$\approx 0.1$], starbursts can rival QSOs as sources of the
metagalactic background at $z > 3$. \cite{Haiman:97} made a
coarse estimate of \fesc for spherical objects of different masses at
various redshift.  They provided a fitting formula, $\log \langle
f_{esc}\rangle = 1.92 \exp[-(z-10)^2/1510]-1$, valid for $z>10$, and
chose a value \fesc$\simeq 1$ for $z<10$.  However, these estimates
are not based on any secure calculation or observation and we do not
reproduce their results.

Theoretical models \citep[hereafter DS94]{Dove:94} of
the radiative transfer of Lyc radiation through disk layers of a
spiral galaxy like the Milky Way suggest that \fesc could range from
5--20\%. The DS94 calculation considered only ``burn-through'':
photoionized channels (H~II regions) produced in smoothly distributed
gas layers in hydrostatic equilibrium.  More realistic models must
include gas dynamics and radiative transfer through a network of
radiatively ionized H~II regions and superbubble-driven cavities
(``chimneys") of hot gas \citep{Norman:89}. By solving the
time-dependent radiation transfer of stellar radiation through
evolving superbubbles, \cite{Dove:99} found that
superbubbles reduce the escape fraction to \fesc $\sim 6\%$ because
the shells of the expanding superbubbles can trap or attenuate the 
ionizing flux. By the time that the superbubbles of large associations 
blow out of the H~I disk, the ionizing photon luminosity has dropped well 
below the maximum luminosity of the OB association.  Observational limits 
on the transmitted flux from four low-redshift galaxies studied by the 
{\it Hopkins Ultraviolet Telescope} (HUT) are consistent with \fesc
ranging from a few percent up to $50\%$.  By comparing the observed
number of Lyc photons with a set of theoretical spectral energy
distributions for the galaxies, \cite{Leitherer:95} derived, on
average, \fesc $\leq 3\%$.  Reexamining the same observations, 
\cite{Hurwitz:97} derived larger upper limits on the escape fraction
($5.2\%$, $11\%$, $57\%$, and $3.2\%$) for the four
galaxies. Absorption from undetected interstellar components could
allow the true escape fractions to exceed these upper limits.
 
In this paper, we attempt to provide a more realistic estimate of the
production rate and escape fraction of Lyc from high-$z$ galaxies. We
estimate the fraction of Lyc flux escaping the halo of spherical
galaxies as a function of their mass, integrated ionizing flux emitted
by the hot star population, and virialization redshift of their dark-matter
halos.  In \S~\ref{sec:meth} we describe the method used to estimate the
escape fraction, and in \S~\ref{sec:res} we show the results of the
simulations. In \S~\ref{sec:ps} we use a Press-Schechter estimate
of the distribution of virialized dark-matter halos to estimate the
emissivity from galaxies as a function of redshift.  Reionization
from the first stars appears to be dominated by small objects with
$M_{DM} \leq 10^{7}~M_{\odot}$. Finally, in
\S~\ref{sec:sum} we summarize and discuss our results.

\section{Method \label{sec:meth}}

We study the Lyc escape fraction from a spherical galaxy as a function
of its virialization redshift, dark matter mass, baryonic fraction and
total Lyc photon flux, $S_{tot}$. We also consider the case of
$S_{tot}$ proportional to the baryonic content of the galaxy, and we
derive from simulations analytical expressions of \fesc for both cases.

We use a Monte-Carlo method to simulate the radial distribution of OB
associations in a spherical galaxy. In our fiducial model (see
\S~\ref{ssec:fm}), we adopt a constant stellar probability
distribution as a function of radius, with a sharp stellar cutoff
at the core radius of the dark matter halo. In \S~\ref{ssec:p} we
relax this assumption, exploring the cases of a stellar density
distribution that follows the baryonic mass distribution and the
Schmidt Law. We also estimate \fesc for the cases when the stars are
located at the center of the halo and when the sharp stellar cutoff is
set to a critical baryonic density, $n_b =1$ and 10 cm$^{-3}$.  We then
calculate $f_{esc}(S)$ for each OB association as a function of its
Lyc photon luminosity, $S$ (photons s$^{-1}$), and its location in the
galaxy. We calculate the number of OB associations with luminosities
$S-\Delta S/2 < S < S+\Delta S/2$ integrating the luminosity function from
the lower to the upper cutoff of the luminosity function, normalized
to the total Lyc photon luminosity. Since we do not use a Monte-Carlo
method but a simple integration, the actual upper cutoff of the
luminosity distribution depends on the total Lyc flux.  In the
fiducial model the luminosity function is $dN_a(S_0)/dS_0 \propto
S_0^{-2}$ for $10^{48} \le S_0 \le 10^{51}$ s$^{-1}$, but in
\S~\ref{ssec:lf} we study the dependence of \fesc on the assumed
luminosity function of the OB associations.  Finally, we sum all
contributions to find \fesc for the whole galaxy, taking care of
averaging the results for several Monte-Carlo realizations of the same
simulation.

The method used to derive the escaping flux for a single OB
association is analogous to that described in DS94, except that we
consider a spherical galaxy instead of a disk galaxy. Thus, we have an
additional degree of freedom, the distance $R$ of the OB associations
from the center of the halo.
We assume that a single OB association is a point source of
Lyc photons and that the gas is in ionization equilibrium.  Since
the hot stars may lie off galactic center, the shape of the 
H II region is not spherical and is determined by equating 
the number of photoionizations with radiative recombinations along
each ray. Along a given ray, the 
Str\"omgren radius, $r_s$, is defined by:
\begin{equation}
    {S_0 \over 4 \pi \alpha_H^{(2)}} = \int_0^{r_s} n_H^2(R)r^2 dr \; ,
\end{equation}
where $S_0$ is the unattenuated Lyc photon luminosity of the OB association
(s$^{-1}$), $n_{H}$ is the hydrogen number density, and
$\alpha_{H}^{(2)}$ is the hydrogen case-B recombination coefficient.
We assume that the gas inside the H~II region is isothermal at
$T=10^4$ K [$\alpha_{H}^{(2)} = 2.59 \times 10^{-13}$ cm$^3$
s$^{-1}$]. The density profile of the baryonic matter is determined by
solving the hydrostatic equilibrium of the gas in the potential well
of the DM halo. We assume a spherically symmetric halo, so
the density dependence is only on $R$. If the OB association lies off
center at a distance $D$, in a polar coordinate system ($r, \theta$)
with the OB association centered at the origin of axes and the
$y$-axis intersecting the center of the halo, we have $R=(D^2+r^2+2Dr
\sin\theta)^{1/2}$. In Figure~\ref{xfig} we draw a sketch showing the
galaxy halo with the system of reference and the notations used in the text.

\def\capxfig{%
Sketch showing the galaxy halo with the system of reference and the
notations used in the text. The OB association has a distance $D$ from
the center of the halo, and the density is a function of the radius
$R$. The critical angle, $\theta_c$, is defined for the ($r, \theta$) 
polar coordinate system centered on the OB association.}
\placefig{
\begin{figure}
\epsscale{1.}
\plotone{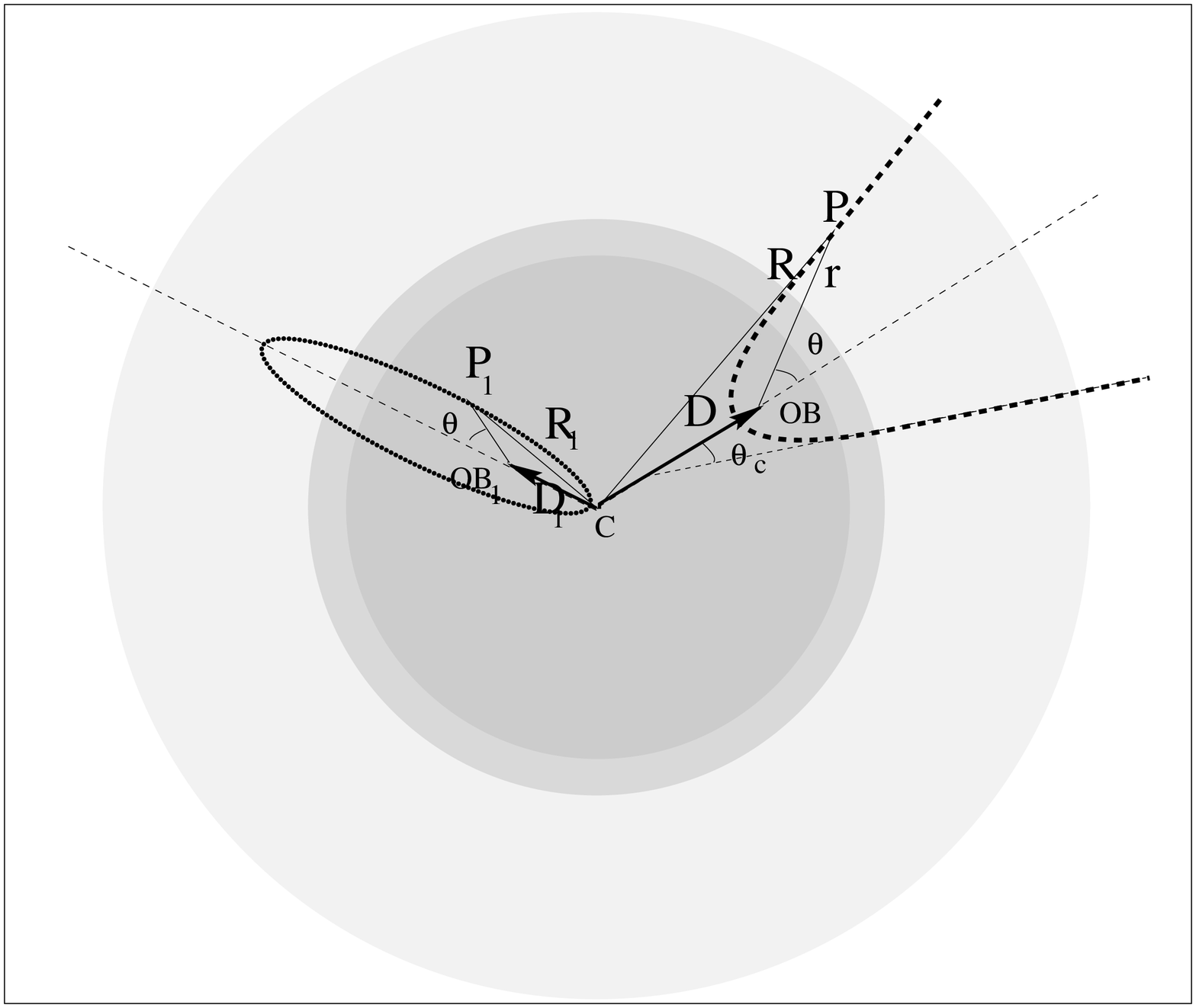}
\caption{\label{xfig}\capxfig}
\end{figure}
} 

The fraction of Lyc photons escaping the halo is given by integrating
over solid angle the fraction $f(\theta)d\theta$ of Lyc
photons emitted between angles $\theta$ and $\theta+d\theta$ that are
not absorbed by the H I in the halo:
\begin{equation}
   \eta_{esc}(S_0) = {1 \over 4\pi}\int_0^{\theta_c(S_0)} f(\theta)2\pi
                   \sin\theta \, d\theta \; ,
\end{equation}
where $\theta_c$ is the critical angle [$f(\theta_c)=0$] for escape
(Figure~\ref{xfig}) and
\begin{equation}
   f(\theta) = 1-{4\pi \alpha_H^{(2)} \over S_0}\int_0^{\infty}
          n_H^2(R)r^2dr \; .
\end{equation}

\subsection{Barionic mass distribution}

The spherical nonlinear model for gravity perturbations predicts that
a just-virialized gas cloud has an overdensity (``collapse factor'')
of $\Delta_c(\Omega_m,\Lambda)=18 \pi^2 \approx 178$ and is
shock heated to the virial temperature. Thus,
the baryonic gas that virializes into the 
DM halo of mass $M_{DM}$ at redshift $z_{vir}$ has average
hydrogen number density $\overline n_{vir}$, virial radius $R_{vir}$
(following convention, this is also called $R_{200}$, where one 
approximates 178 by 200), and temperature $T_{vir}$ given by:
\begin{eqnarray}
\overline n_{vir} &\sim& 0.03~{\rm cm}^{-3}\left({\Omega_b h^2 \over
     0.019}\right)\left({1+z_{vir} \over 10}\right)^3,\\
R_{vir} &\sim& 344~{\rm pc}\left({\Omega_m h^2 \over 0.15}
    \right)^{-1/3}\left({M_{DM}
     \over 10^6~{\rm M}_{\odot}}\right)^{1/3}\left({1+z_{vir} \over
     10}\right)^{-1},\label{eq:r200}\\
T_{vir} &\sim& 556~{\rm K}\left({\Omega_m h^2 \over 0.15}\right)^{1/3}\left({M_{DM}
    \over 10^6~{\rm M}_{\odot}}\right)^{2/3}\left({1+z_{vir} \over
    10}\right)\left({\mu \over 0.6 m_p}\right),
\end{eqnarray}
where $\Omega_m$, $\Omega_b$ and $h$ have the usual meanings and $\mu$
is the mean molecular weight of the gas.
 
We assume a hierarchical clustering scenario for the formation of
virialized halos \citep{White:91}. The results of numerical simulations 
show that DM halos are well fitted by a universal profile
of mass density, $\rho(R)$, valid for a wide range of halo masses.  According 
to the results of \cite{Navarro:96, Navarro:97} this profile is 
steeper than that of an isothermal sphere, $\rho(R) \propto R^{-3}$ 
if $R>R_s$ and smoother if $R<R_s$. Thus, we take
\begin{equation}
   \rho_{DM}(R) = {\delta_c ~ \rho_{c0} \over (R/R_s)(1+R/R_s)^2} \; ,
\end{equation}
where $\rho_{c0}=3H_0^2/8\pi G$ is the critical density of the
universe at $z=0$.  Integrating $\rho_{DM}(R)$, one finds that the overdensity
is given by
\begin{equation}
   \delta_c = (3 \times 10^3)  \Omega_m (1+z_{vir})^3 \; .
\end{equation}
The characteristic radius, $R_s(M)$, of the DM is given by 
\begin{equation}
   R_s(M)={R_{vir}(M_{DM}) \over c(M_{DM})}={1 \over
   c(M_{DM})}\left({3M_{DM} \over 4 \pi \Delta_c 
         \rho_{c0}}\right)^{1 \over 3} \; ,
\end{equation}
where $c$ is the concentration factor, $\Omega_m$ is the 
density parameter at $z=0$, $\Delta_c(\Omega_m,\Lambda)$ is the 
collapse factor in a spherical nonlinear model, and $z_{vir}$ is the 
virialization redshift of the halo. The
concentration parameter $c$ is related to $\delta_c$ by
\begin{equation}
   \delta_c = \Delta_c \left[ {c^3 \over \ln(1+c)-c/(1+c)} \right] \; . 
\end{equation}
We solve the equation of hydrostatic equilibrium of the baryonic
matter numerically by assuming that the gravitational effect of the
baryonic matter is negligible compared to the DM.  The support
of the gas in the gravitational well of the DM is dominated by bulk
motions at high redshifts \citep{AbelAnninos:98}. We then define
an effective temperature profile that is the sum of all contributions
to the pressure (turbulent pressure, ram pressure, centrifugal force).
%We assume an {\it a priori}
%effective temperature profile for the baryonic matter. 
If the gas is
isothermal, the equations have a simple analytical solution,
\begin{equation}
    \rho_g(R) = \rho_{g0} \exp (-27b/2) \left(1+{R \over R_s}\right)^{27bR_s 
                                          \over 2R} \; ,
\end{equation}
where $b=(2c/9\Gamma)[\ln(1+c)-c/(1+c)]^{-1}$ and where
$\Gamma=T_{vir}/T_{g} \simeq 1$ measures the efficiency of shock
heating of the gas. If we assume isothermality, the resulting profile
is well approximated by a $\beta$-model \citep{Makino:98},
\begin{equation}
\rho_{g}(R)={\rho_{g0} \over [1+(R/R_c)^2]^{3\beta/2}} \; ,
\label{eq:betam}
\end{equation}
with $\beta \simeq 0.9b$ and $R_c \simeq 0.22 R_s$. In
Figure~\ref{density} we show the DM density profile and the baryonic
matter density profile for an isothermal gas.

\def\capden{%
Profile of baryonic matter (BM, solid line) expected
from the universal density profile of DM halo (dotted line)
for an isothermal gas at virial temperature $T$ (dashed line). The
radius $R_s=R_{200}/c$ is related to the size of the DM core,
where $R_{200}$ is the virial radius and $c$ is the concentration
parameter; in this plot $c=10$. The density profile of the gas is well
approximated by the isothermal $\beta$-model with a core radius
$R_c=0.22R_s$ and $\beta=0.9b$ ($b \sim 1$ is a function of the
concentration parameter and the gas temperature).}
\placefig{
\begin{figure}
\epsscale{1.}
\plotone{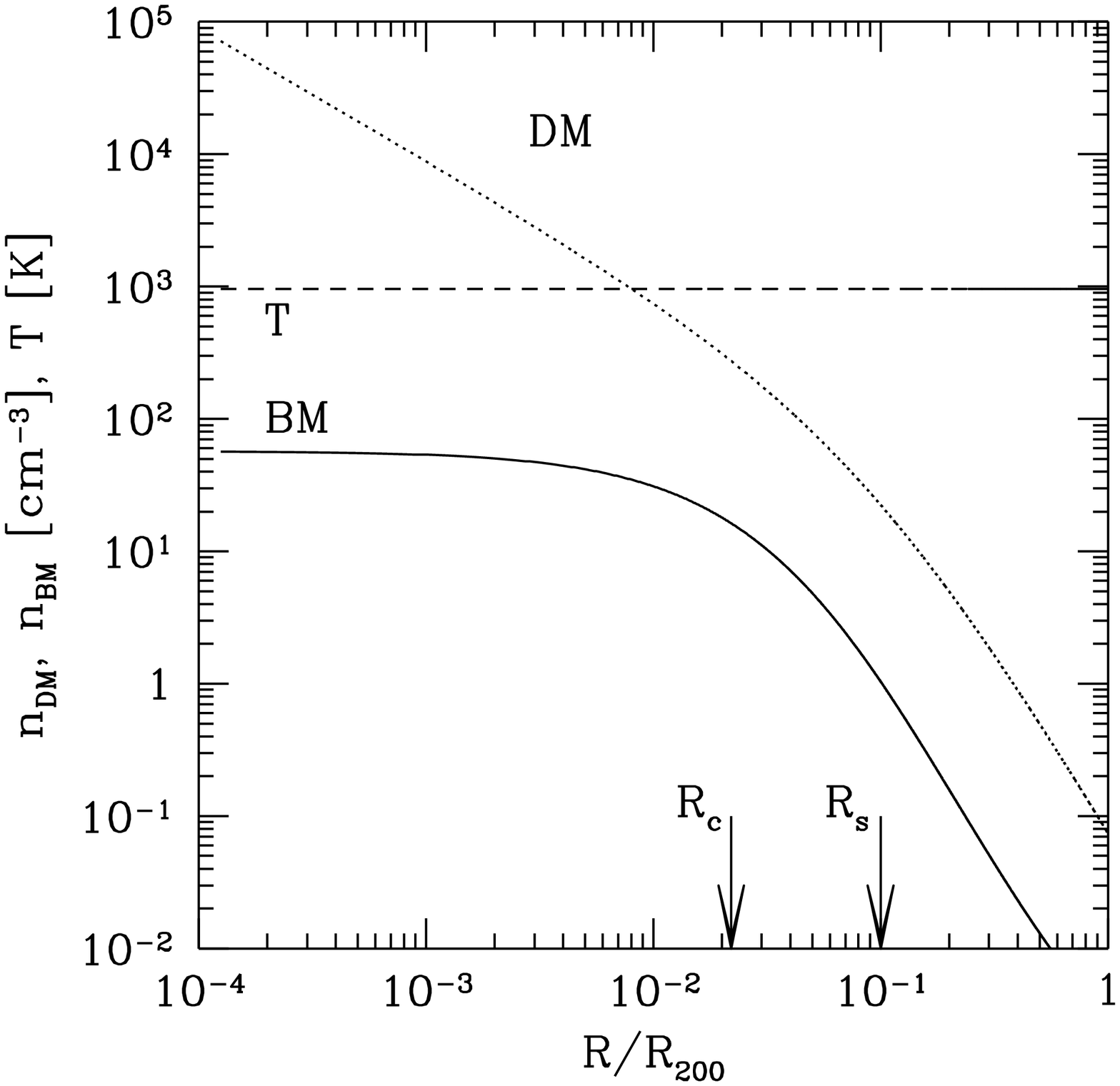}
\caption{\label{density}\capden}
\end{figure}
}

\subsection{Stellar density distribution and luminosity function \label{ssec:fm}}

We estimate the total number of escaping Lyc photons from the galaxy by integrating the escaping radiation of each single OB association
over the luminosity function of the OB associations, where the
position of each association inside the halo is determined by a
Monte-Carlo simulation. We assume a power law for the luminosity
function, $dN_a(S_0)/dS_0 \propto S_0^{-\alpha}$ for $S_1 \le S_0 \le
S_2$, where $S_1$ and $S_2$ are lower and upper limits of the observed
luminosities and $\alpha = 2.0 \pm 0.5$, as has been determined for OB
associations in 30 local spiral and irregular galaxies
\citep{Kennicutt:89}. A value $\alpha = 2.12 \pm 0.04$ was also 
found for the luminosity function of young stellar clusters in the
interacting Antennae galaxies \citep{Whitmore:99}.   
In our fiducial models, we assume $S_1=10^{48}$
s$^{-1}$, $S_2=10^{51}$ s$^{-1}$ and $\alpha=2$; for a more exhaustive
discussion of this assumption, see DS94.

The flux emitted by all the OB associations is given by:
\begin{equation} 
   S_{tot} = \int_{S_1}^{S_2} S_0\left({dN_a(S_0) \over dS_0}\right) dS_0 =
                  N_{OB}S_1\ln\left({S_2 \over S_1}\right)=N_{OB}(6.9
                  \times 10^{48}~{\rm s}^{-1}) \; ,
\end{equation}
where $N_{OB}$ is the number of OB associations. We divide the OB
associations into 10 logarithmic bins according to their
luminosity. In the case of $\alpha=2$ and 10 logarithmic bins, the
upper cutoff is smaller than $S_2$ if $S_{tot} < 10 \times S_2$.

We assume a radial density distribution for the associations.  In the
fiducial models we chose a random distribution in radius for the
distance of the associations from the center of the halo with a cutoff
at a radius $R=R_s$.  In this region, if the gas is isothermal, the
density profile is almost flat inside the core radius $R=R_c$ and
starts to decrease near $R_s$ as $R^{-3}$. The average OB association
density distribution then goes as $R^{-2}$ for $R=0$ to $R=R_s$ and is
zero for $R>R_s$. In \S~\ref{ssec:lf} we study the dependence of \fesc
on the assumed luminosity function of the OB associations, and in
\S~\ref{ssec:p} we study the effect on \fesc of different stellar
density distributions and radial cutoffs.

\def\capsim{%
Shapes of H II regions in one Monte-Carlo
simulation with $N_{OB}=100$ ($S_{tot}=6.9 \times
10^{50}$ photons s$^{-1}$) in a DM halo with $M_{DM}=10^9$~M$_\odot$,
$R_s \simeq 1.1$ kpc (outer circle), and $z_{vir}=2$. The gas core radius is
$R_c \simeq 250$ pc (inner circle), and the gas effective temperature
is $T \simeq 16,000$ K. In this simulation we find \fesc$\simeq 6.5\%$.}
\placefig{
\begin{figure}
\epsscale{1.}
\plotone{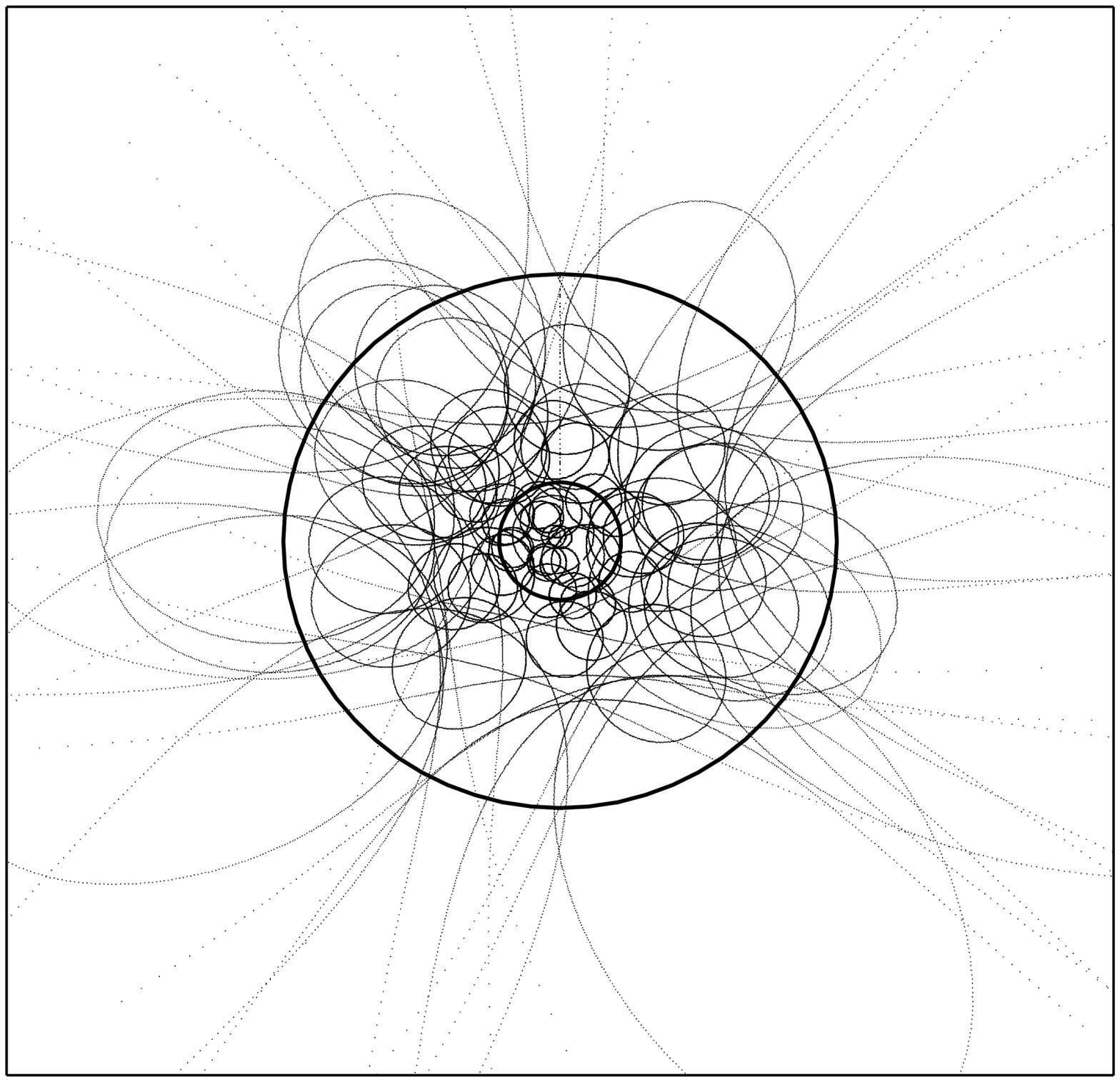}
\caption{\label{simulation}\capsim}
\end{figure}
}

\section{Results\label{sec:res}}

The cosmological model we adopt is $\Lambda$CDM with $\Omega_m=0.3$,
$\Omega_\Lambda=0.7$, $h=0.7$ and $\Omega_b h^2=0.019$ (Burles \&
Tytler 1998). Unless otherwise stated, the DM halos are characterized
by a concentration parameter $c=10$ and a collapse factor 
$\Delta_c(\Omega_m,\Lambda) \approx 200$. We do not consider a
multi-phase interstellar medium, and we neglect the effect of stellar
winds and supernova explosions on the dynamics of the ISM \citep{Dove:99}.
Later in this section, we discuss these approximations.

\subsection{Case I: constant Lyc flux}

First, we consider a constant number of OB associations (\ie, fixed
$S_{tot}$), in galaxies of different masses. An example of a
Monte-Carlo simulation is shown in Figure~\ref{simulation}, and our
results are shown in Figure~\ref{fig:mz}.  In Figure~\ref{fig:mz} (left) we show the
escaping fraction as a function of the mass of the DM halo for a
constant starburst that produces a total number of Lyc photons
$S_{tot}=3.5 \times 10^{51}$~s$^{-1}$. Each point is the mean of five
Monte-Carlo simulations with identical parameters; the error bars show
the variance of the mean and the three curves refer to different
virialization redshifts. In Figure~\ref{fig:mz} (right), we show, for a fixed
redshift $z_{vir}=10$, the dependence of the escaping fraction on both
$S_{tot}$ and the DM mass of the halo. The upper cutoff of the
luminosity distribution is $S_{up}=min(S_{tot}/10,S_2)$; therefore if
$S_{tot}>10^{52}$ s$^{-1}$ it remains constant. From
Figure~\ref{fig:mz} (right) we notice that \fesc remains constant if
$S_{tot}>10^{52}$ s$^{-1}$ indicating that \fesc depends on $S_{up}$
but not on $S_{tot}$. A good fit to the points in Figure~\ref{fig:mz} is
given by a power-law function of the mass with an exponential cutoff
at the critical mass $M^*$,
\begin{equation}
\langle f_{esc} \rangle = N \left({M_{DM}\over M^*}\right)^{-\beta} \exp \left[-\left({M_{DM}\over
M^*}\right)^2\right].
\label{eq:fit1}
\end{equation}
The lines in Figure~\ref{fig:mz} are the best fits to the points, using
eq.~(\ref{eq:fit1}) with the fitting parameters $N$, $\beta$ and $M^*$
listed in Table~\ref{tab:1}. We remind the reader that this first
expression for \fesc is for the case $S_{tot}=const$.

\def\tabone{
\begin{deluxetable}{lccccc}
%\footnotesize
\tablecaption{Best-fit parameters (see eq.~\ref{eq:fit1}) for the
curves in Figure~\ref{fig:mz} and Figure~\ref{fig:cut}.\label{tab:1}}
\tablewidth{0pt}
\tablehead{
\colhead{$z_{vir}$} & \colhead{$S_{tot}$ (s$^{-1}$)} &
\colhead{$R_{cut}/R_s$} & \colhead{$N$} & \colhead{$\beta$} &
\colhead{$\log M^*$ (M$_\odot$)}
 }
 \startdata
 3  & $3.5 \times 10^{51}$ & 1 & 2.0\% & 0.4 & 9.5 \\
 5  & $3.5 \times 10^{51}$ & 1 & 0.8\% & 0.5 & 9.3 \\
 10 & $3.5 \times 10^{51}$ & 1 & 0.3\% & 0.65 & 8.8 \\
 10 & $3.5 \times 10^{52}$ & 1 & 1.4\% & 0.4 & 9.1 \\
 10 & $1.0 \times 10^{53}$ & 1 & 1.6\% & 0.4 & 9.2 \\
 15 & $3.5 \times 10^{51}$ & 1 & 0.4\% & 0.8 & 7.8 \\
 10 & $3.5 \times 10^{52}$ & 5 & 2.7\% & 0.4 & 8.7 \\
 10 & $3.5 \times 10^{52}$ & 0.5 & 5.4\% & 0.3 & 9.3 \\
 10 & $3.5 \times 10^{52}$ & 0.1 & 17.0\% & 0.1 & 10.5 \\
 \enddata
\end{deluxetable}
}
\placefig{\tabone}
\def\capfmz{%
  (left) Escaping fraction of Lyc photons as a function of the mass of
  the DM halo for a constant starburst that produces $S_{tot}=3.5
  \times 10^{51}$ s$^{-1}$. Each point is the mean of five Monte-Carlo
  simulations with identical parameters; the error bars show the
  variance of the mean. The four curves refer to virialization
  redshifts from $z_{vir}=3$ to 15. (right) Same quantity for a constant
  starburst at $z=10$. The three curves refer to different total
  number of Lyc photons: $S_{tot}=10^{53}$, $3.5 \times 10^{52}$, and
  $3.5 \times 10^{51}$ s$^{-1}$. We note that \fesc does not depend on
  $S_{tot}$ if $S_{tot} > 10^{52}$ s$^{-1}$. Some points at low mass
  and high total number of emitted Lyc photons shown in the figure are
  unrealistic, but we show them for illustrative reasons.}  
\placefig{
\begin{figure}
\epsscale{1.1}
\plottwo{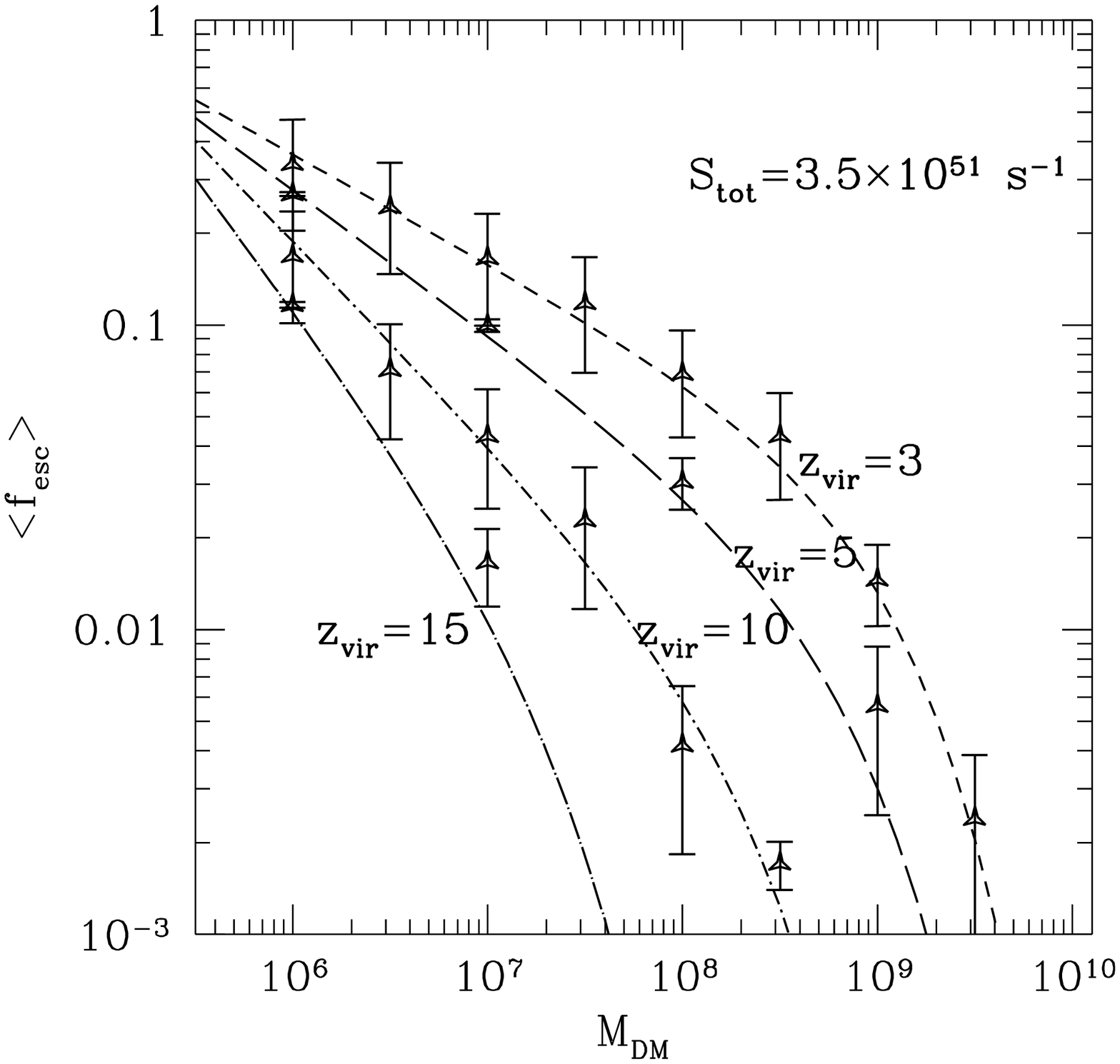}{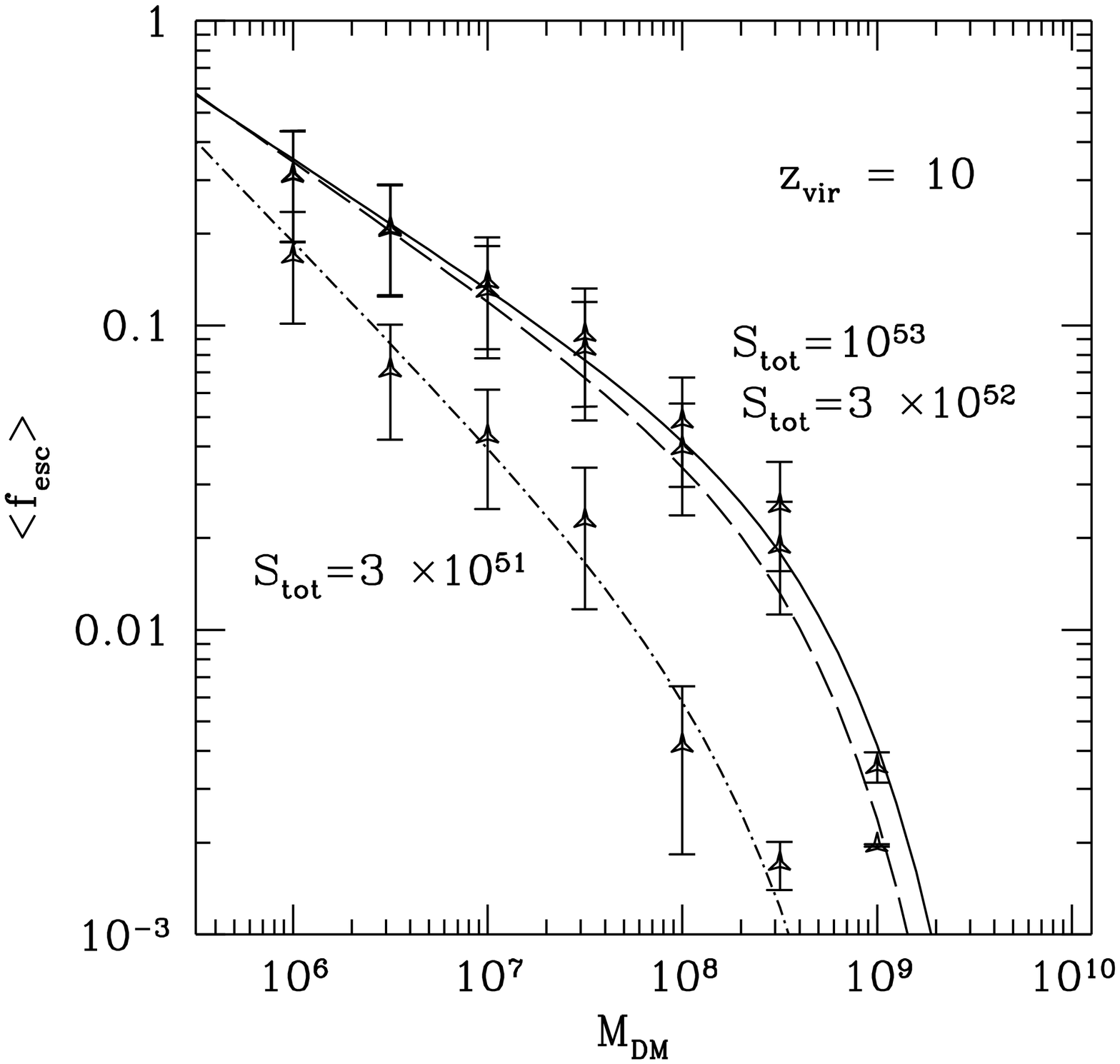}
\caption{\label{fig:mz}\capfmz}
\end{figure}
}
\def\capcut{%
Escape fraction as a function of the dark-matter mass for a galaxy at
$z_{vir}=10$ with $S_{tot}=3.5 \times 10^{52}$~
s$^{-1}$. Different curves refer to situations where, outside a fixed
radius cutoff $R_{cut}$, the gas is completely ionized by collisional
ionization.}
\placefig{
\begin{figure}
\epsscale{1.}
\plotone{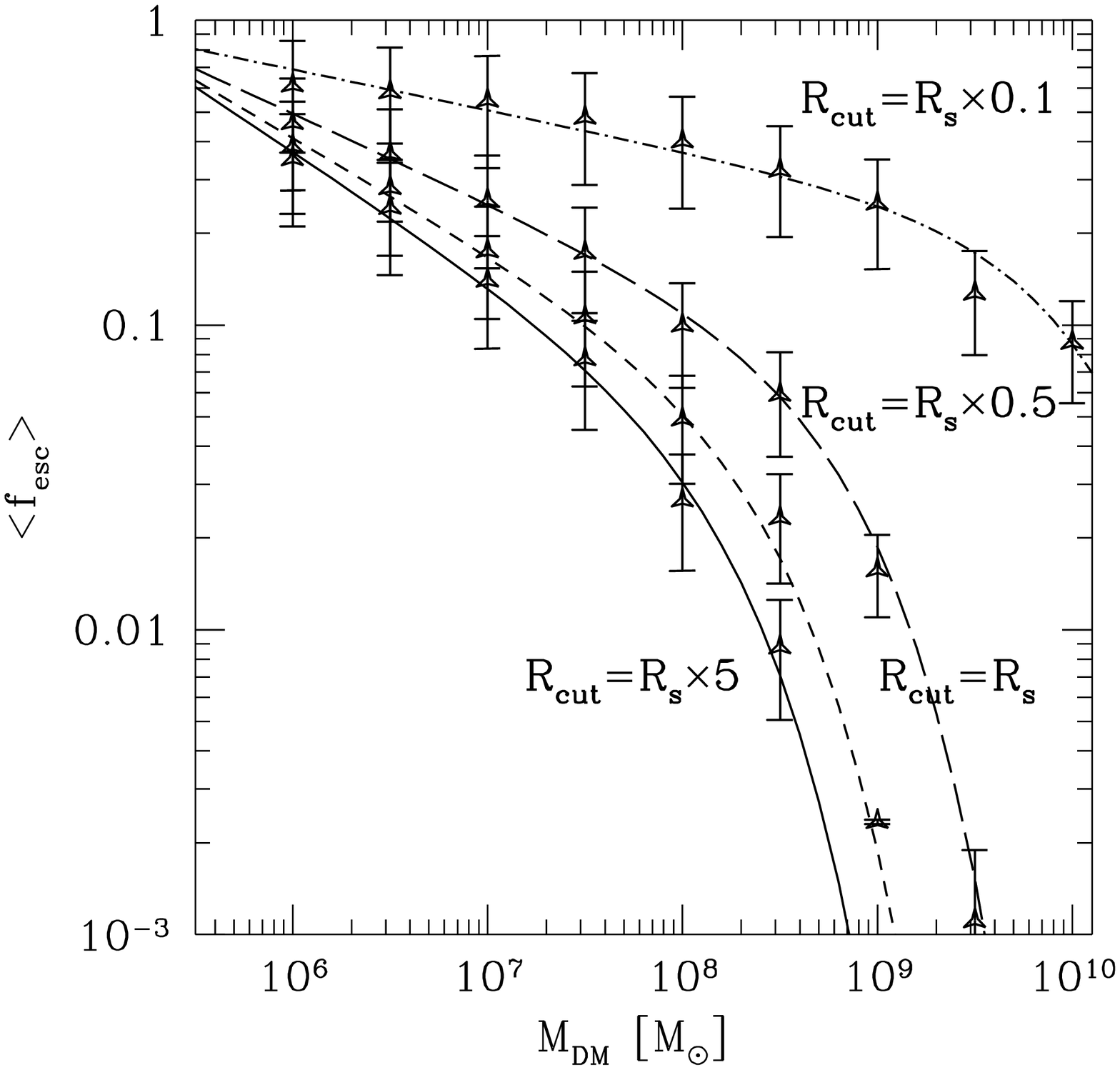}
\caption{\label{fig:cut}\capcut}
\end{figure}
}

If the DM halo is more massive than $M_{ion} \sim 6.8 \times
10^9$~M$_\odot(1+z_{vir})^{-1.5}(\Omega_m h^2/0.15)^{-1/2}$, it is
likely that the gas falling into the potential well is collisionally
ionized by the shock waves that virialize the gas. This effect is more
likely in the outer parts of the halo. When the ionizing photons from
an OB association reach the radius where the gas is collisionally
ionized ($T \simgt 2 \times 10^4$~K) we assume that they are free to
escape. Thus, the escaping fraction increases with decreasing values
of the inner radius where the gas starts to be collisionally ionized.
We show this result in Figure~\ref{fig:cut} for the case $z_{vir}=10$
and $S_{tot}=3.5 \times 10^{51}$ s$^{-1}$. The curves are a parametric
fit to the points, using eq.~(\ref{eq:fit1}) with fitting parameters
listed in Table~\ref{tab:1}.

\subsection{Case II: Lyc flux proportional to the mass}

\def\capfz{%
  Results of fiducial models.  Escaping fraction of Lyc as a function
  of the virialization redshift assuming a total Lyc flux, $S_{tot} =
  (1.14 \times 10^{49}~{\rm s}^{-1}) \epsilon
  f_g(M_{DM}/10^6~M_\odot)$, proportional to the mass of the galaxy,
  $M_g = f_g(\Omega_b / \Omega_m) M_{DM}$. The error bars are the
  variance of the mean on several Monte-Carlo simulations, the dashed
  lines are linear fits to the mean values of \fesc and the solid
  lines the analytical formula eq.~(\ref{eq:fit}). (top-left) Here we
  assume $\epsilon=4$ (\ie, 4 times the Milky Way SFE), collapsed gas
  fraction $f_g=0.5$ and different curves are relative to $M_{DM}=
  10^7, 10^8, 10^9$~M$_\odot$. (top-right) Assumes $f_g=1, \epsilon=4$
  and $M_{DM}=10^6, 10^7, 10^8, 10^9$~M$_\odot$. (bottom-left) Same as
  the top-right panel but for $\epsilon=40$. (bottom-right) Same as
  the top-right panel but for $\epsilon=400$.}
\placefig{
\begin{figure}
\epsscale {1.1}
\plottwo{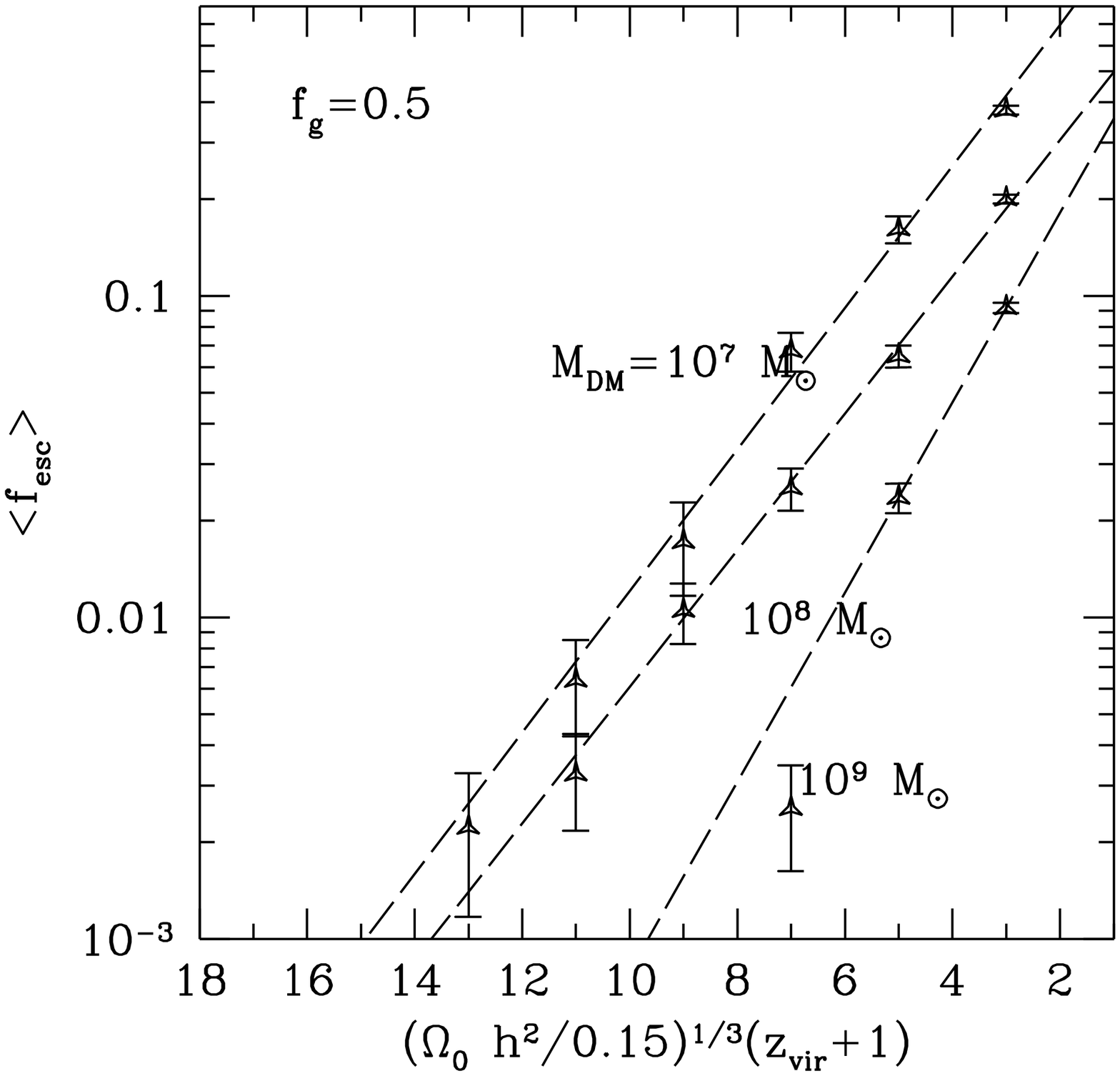}{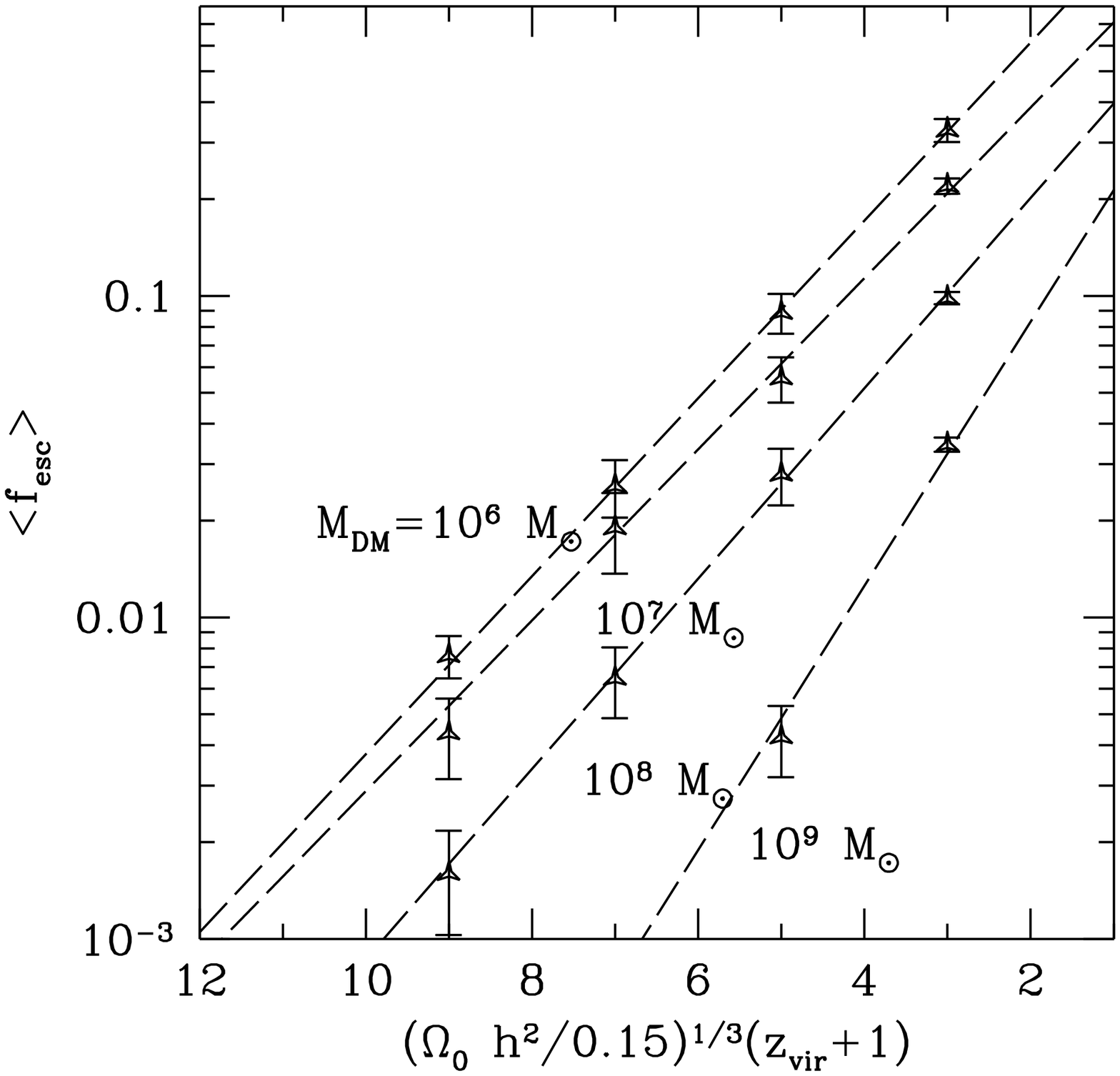}
\epsscale {2.45}
\plottwo{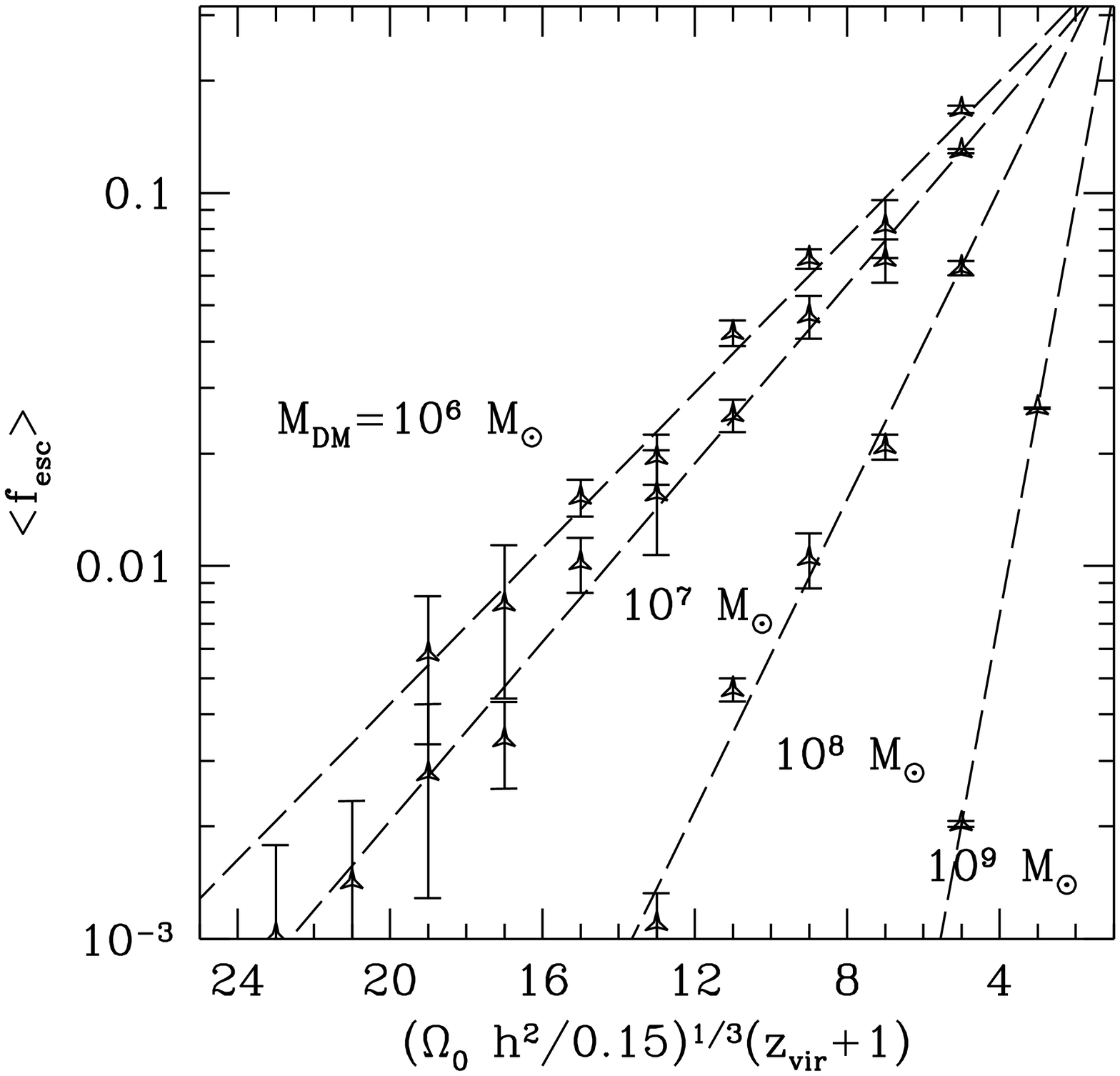}{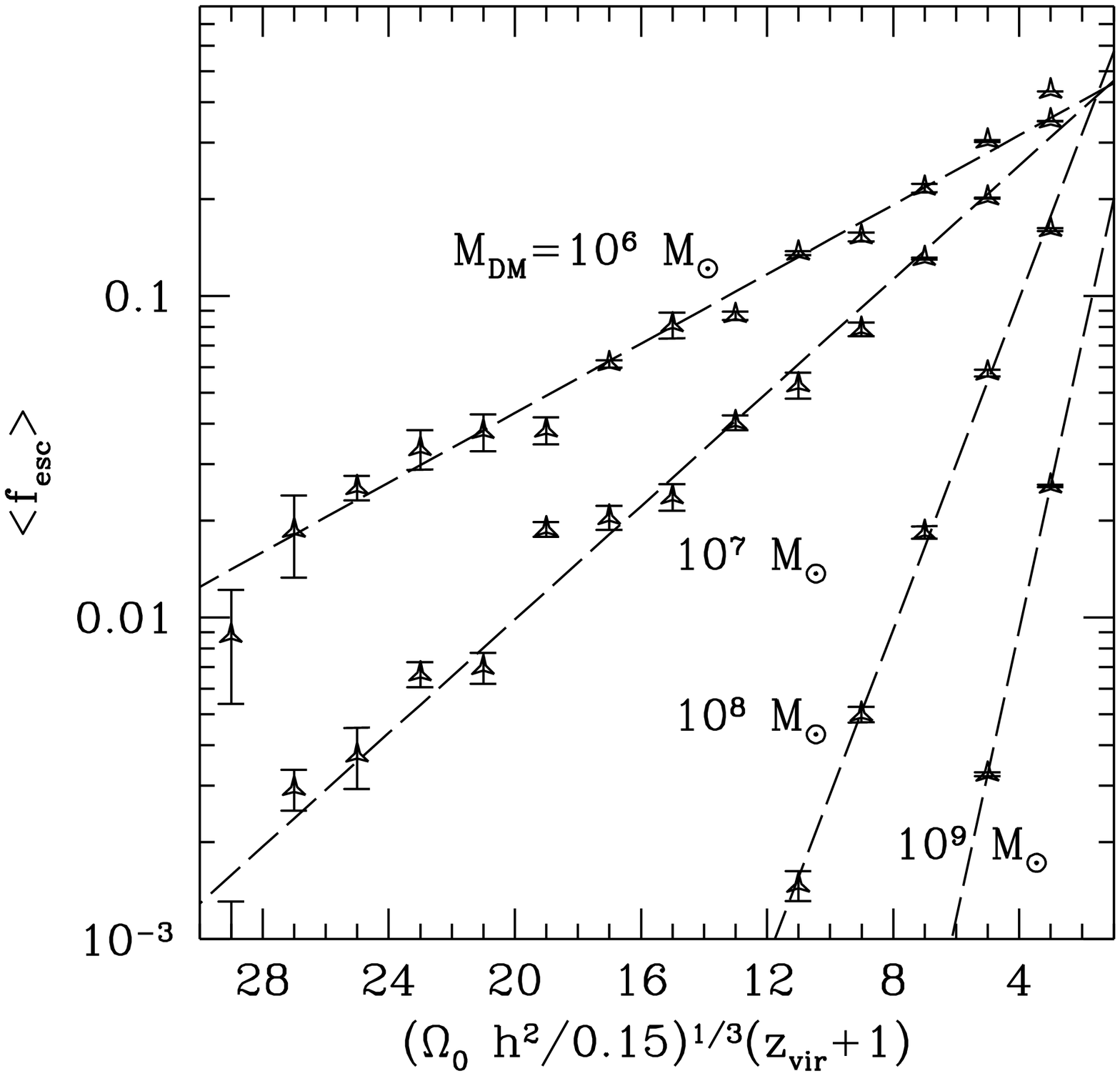}
\caption{\label{f_z1}\capfz}
\end{figure}
}

\def\capf5{%
  (left) Escaping fraction of Lyc as a function of the mass of the halo
  for $\epsilon=4$ and $S_{tot}\propto \epsilon M_b$. The solid lines
  are the analytical formula, eq.~(\ref{eq:fit}), at
  different virialization redshifts. Note the abrupt change in the
  shape of the escaping fraction for $S_{tot} \geq 10^{52}$
  s$^{-1}$ when the maximum luminosities of OB associations reach a
  sharp cutoff. This behavior remain true also for different SFEs. (right)
  Escaping fraction as a function of the virialization redshift for
  $\epsilon=4, M=10^7$ M$_\odot $ and gas collapsed fraction $f_g=0.25, 0.5, 0.75, 1$. The dashed
  lines are the best fit and the solid lines the analytical formula
  eq.~(\ref{eq:fit}).}
\placefig{
\begin{figure}
\epsscale{1.1}
\plottwo{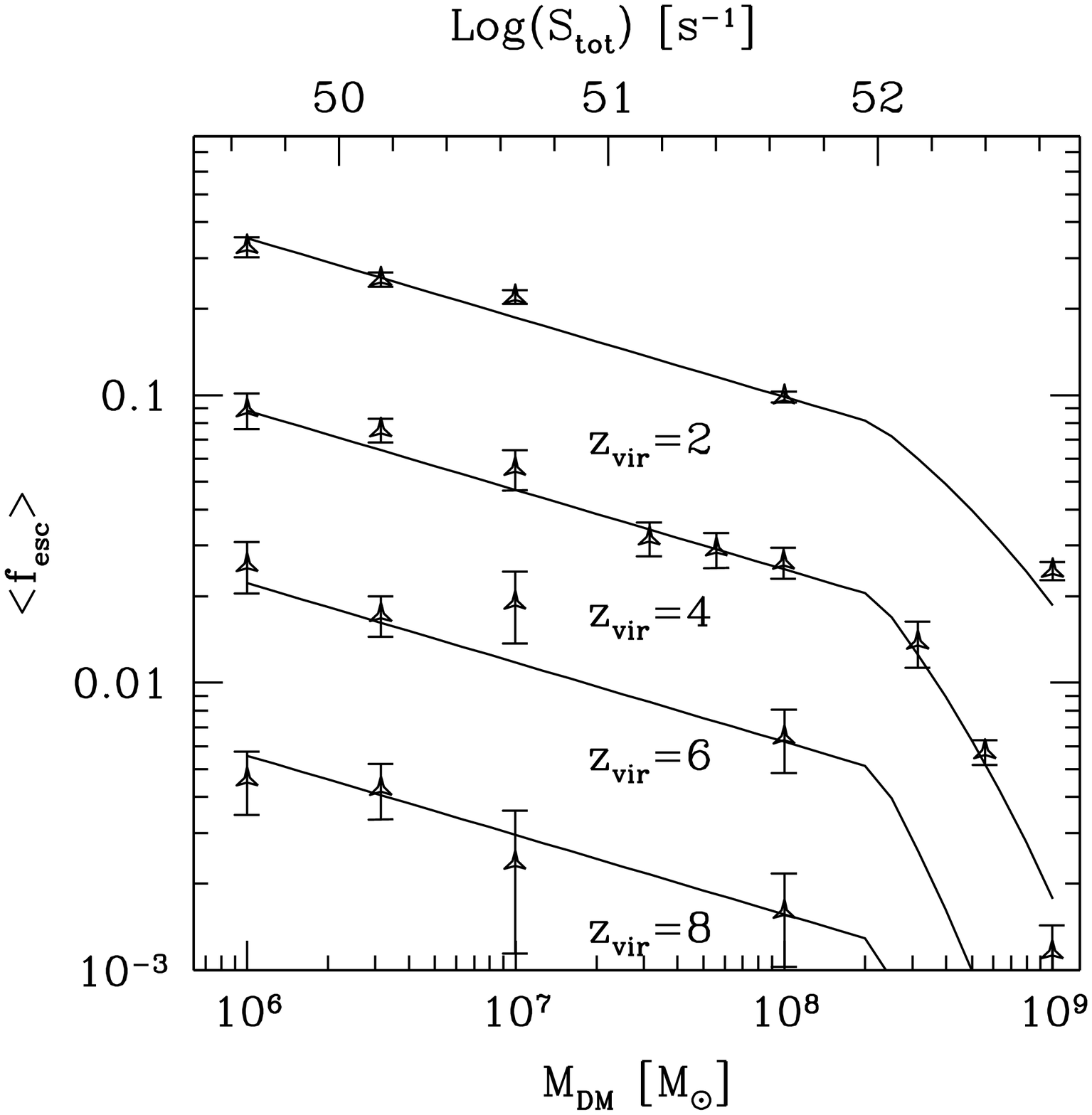}{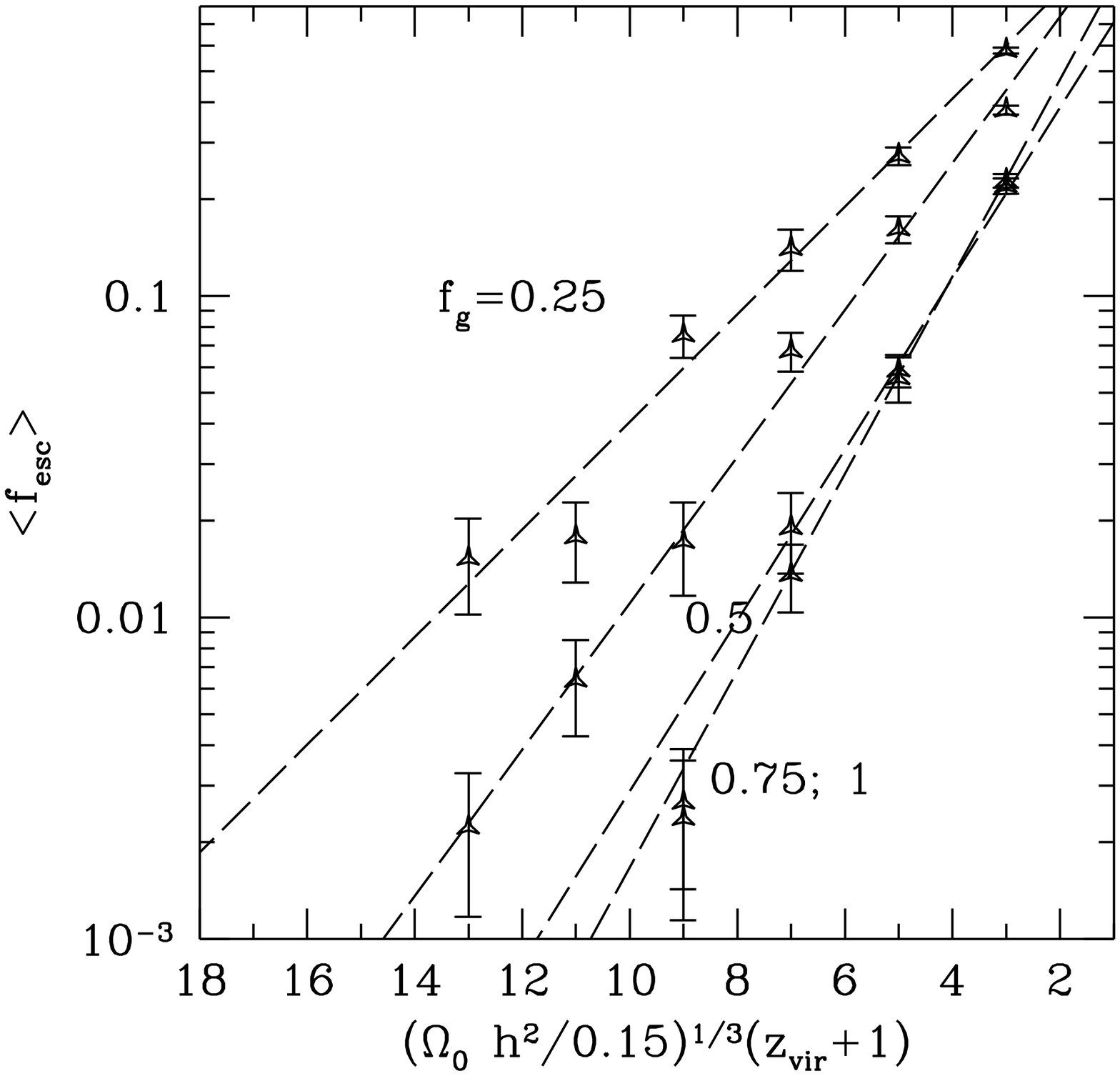}
\caption{\label{fig:5}\capf5}
\end{figure}
}

As second case, we scale the total ionizing flux, and therefore the
number of OB associations in the galaxy, by the linear relation
$S_{tot} = \epsilon \gamma^*(M_{g}/M_{g}^*) \sim (8.75 \times
10^{49}~{\rm s}^{-1}) \epsilon (M_{g}/10^6$~M$_\odot)$, where $M_{g}^*
\simeq 4 \times 10^9$~M$_\odot$ is the gas mass in the Milky Way and
$\gamma^*=3.5 \times 10^{53}$ s$^{-1}$ \citep{Bennett:94}.  The free
parameter $\epsilon$ expresses the SFE normalized to the Milky Way
value.  Our adopted range ($4<\epsilon<400$) represents values of SFE
for which appreciable Lyc escape occurs, in starbursts with rates much
higher than that of the Milky Way.  With the assumed baryonic
fraction, $M_g = f_g(\Omega_b/\Omega_m)M_{DM}\sim 0.13f_gM_{DM}$,
where $f_g \simlt 1$ is the product of the baryonic collapsed fraction
and the fraction of baryons in the form of gas, $S_{tot}$ scales with the
DM mass of the halo as $S_{tot} = (1.14 \times 10^{49}~{\rm s}^{-1})
\epsilon f_g(M_{DM}/10^6$~M$_\odot)$.
 
In Figure~\ref{f_z1} we show \fesc for our ``fiducial models'' as a
function of the virialization redshift of the galaxies and their DM
content for three different values of the parameter $\epsilon$ with
$f_g=1$ and for $\epsilon=4, f_g=0.5$. Each point is the mean of five
Monte-Carlo simulations with identical parameters; the error bars are
the variance of the mean. In a log-linear plot \fesc is, with good
approximation, a linear function of the redshift.  In
Figure~\ref{fig:5} (left) we show the logarithm of the escaping fraction as a
function of the total ionizing flux $S_{tot}$ (proportional to the
mass $M_{DM}$) for different virialization redshifts at a fixed SFE
($\epsilon=4$) and in Figure~\ref{fig:5} (right) we show \fesc for $\epsilon=4$
and $M_{DM}=10^7$ M$_\odot$ for $f_g=0.25, 0.5, 0.75, 1$. We also show
(solid lines) an analytical formula,
\begin{equation}
 \log \langle f_{esc} \rangle = 
\begin{cases}
\log \left({S_{tot} \over 10^3S_1}\right)^{-{1 \over
   3}}-0.41(1+\log f_g^{2 \over 3})(z_{vir}+1)\left({\Omega_m h^2
   \over 0.15}\right)^{1 \over 3}\epsilon^{-{1 \over
   3}}, & \text{if $S_{tot}<10S_2$}\\
\log \left({S_2 \over 10^2S_1}\right)^{-{1 \over
   3}}-0.41(1+\log f_g^{2 \over 3})(z_{vir}+1)\left({\Omega_m h^2
   \over 0.15}\right)^{1 \over 3}\left[\left({10^{51}~{\rm s}^{-1}
   \over S_2}\right){f_g M_{DM} \over 8.8 \times
   10^8~{\rm M}_\odot}\right]^{1 \over 3},& \text{if $S_{tot}>10S_2$}
\end{cases}
\label{eq:fit}
\end{equation}
that is a good fit to the points in Figures.~\ref{f_z1} and \ref{fig:5} over
the range of SFEs ($4<\epsilon<400$), masses, and redshifts shown in
the figures. In the summary we write the same formula in a more
compact form.

The fitting formula has been derived from all the simulation results,
some of them not shown in the paper for sake of brevity. Looking at
Figure~\ref{fig:5} (left), we note the abrupt change in the shape of \fesc
for $S_{tot}>10^{52}$ s$^{-1}$, when the upper cutoff of the
luminosity function stays constant. If we do not set an upper cutoff to
the luminosity function, the dependence of \fesc on the mass of the
halo is a $1/3$ power-law. Instead, if we set the upper cutoff, \fesc drops
exponentially with the mass of the halo and \fesc depends only on the
mass and virialization redshift.
In conclusion, small objects have bigger \fesc than the massive ones with
the same SFE.  

\subsection{Changing the Stellar Density Distribution \label{ssec:p}}

\def\cappa{%
  Escaping fraction of Lyc as a function of redshift for different
  choices of the stellar density distribution; the halo mass is always
  $M_{DM}=10^6$ M$_\odot$. (left) Dot-dashed lines show \fesc when all
  the stars are in the center of the halo for $\epsilon=4,40$. The
  long-dashed lines show \fesc when $\rho^{\ast} \propto \rho_b$ with
  stellar cutoff at $R(n_b=1~{\rm cm}^{-3})$ for $\epsilon=4,40$.
  The dashed line has stellar cutoff at $R(n_b=10~{\rm cm}^{-3})$
  and $\epsilon=4$. Our fiducial models for $\epsilon=4,40$ are shown
  for comparison (solid lines).  (right) The dot-dashed line shows \fesc
  when $\rho^* \propto \rho_b$ with stellar cutoff at $R=R_s$ and
  $\epsilon=4$. The long-dashed lines show \fesc when $\rho^* \propto
  \rho_b^2$ with stellar cutoff at $R(n_b=1~{\rm cm}^{-3})$ for
  $\epsilon=4,40$.  The other lines, shown for comparison, are the
  same as in the left panel.}  
\placefig{
\begin{figure}
\epsscale{1.1}
\plottwo{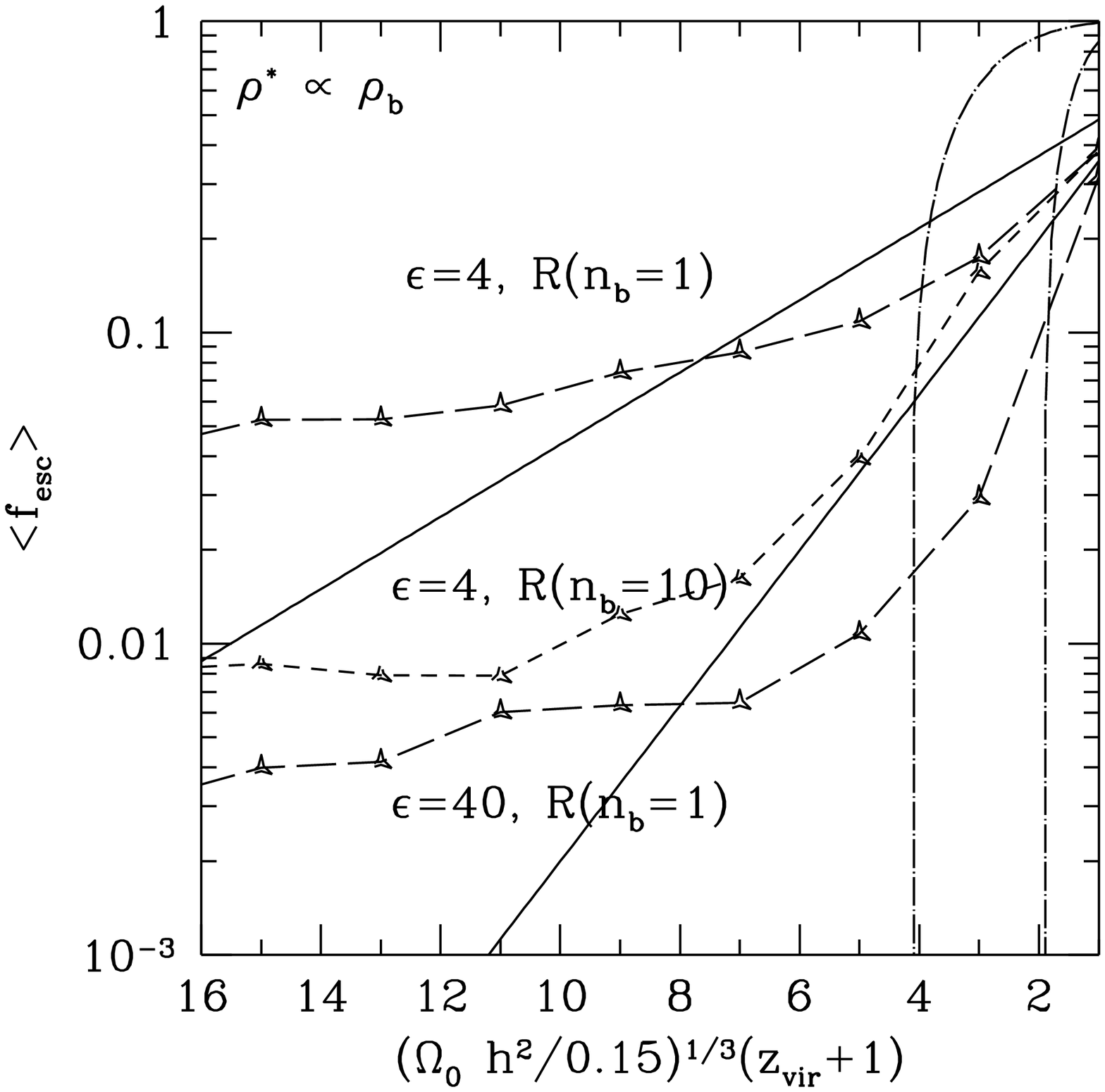}{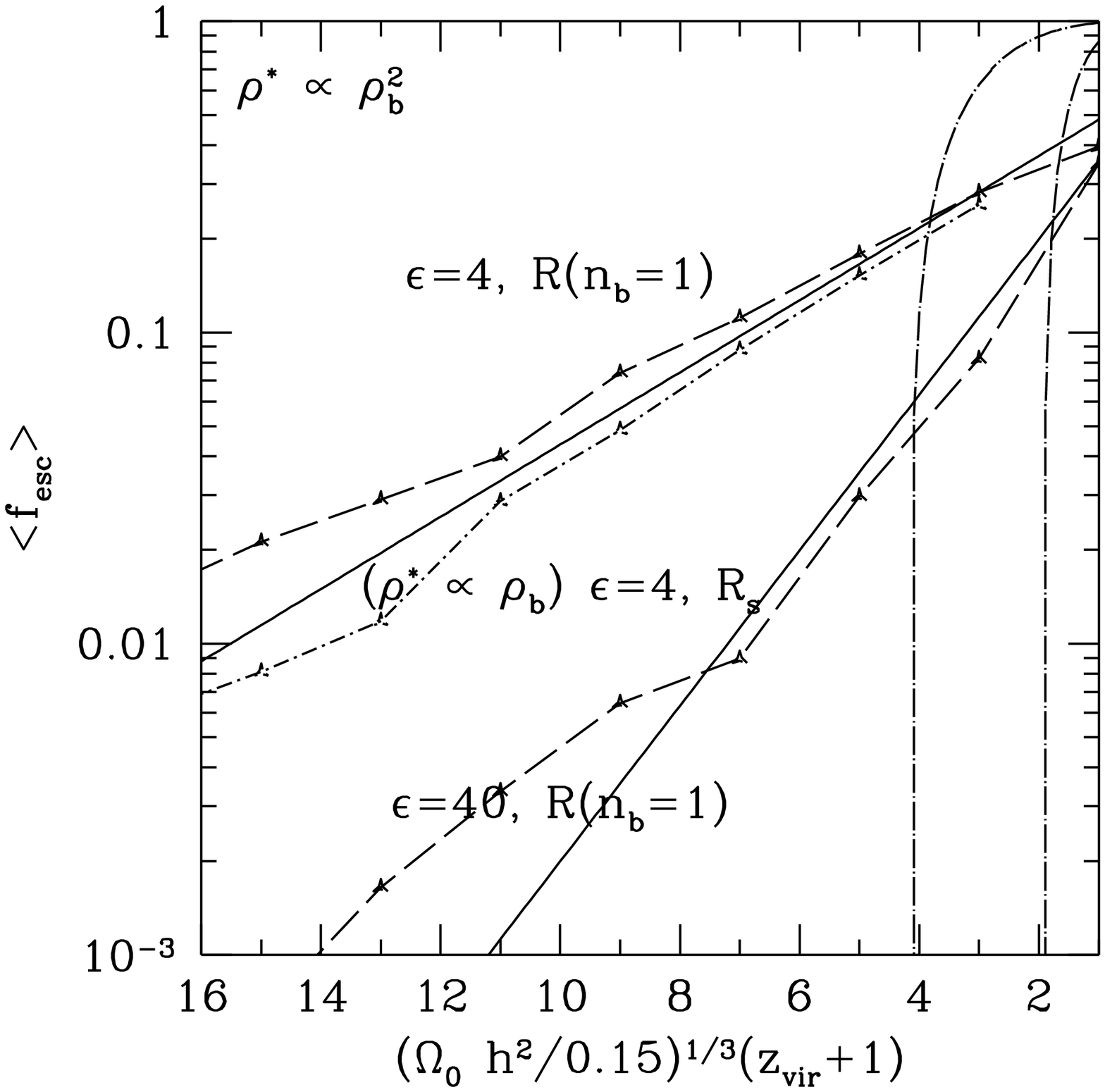}
\caption{\label{fig:p1}\cappa}
\end{figure}
}

The density distribution of massive stars in the halo is crucial for
determining \fesc. If all the stars are located at the center of the
DM potential, it is trivial to find \fesc for a given baryonic density
profile, because the calculation is reduced to the case of a single
Str\"omgren sphere,
\begin{equation}
   f_{esc} = 1-{4\pi \alpha_H^{(2)} \over S_{tot}}\int_0^{\infty}
          n_H^2(R)r^2dr \;.
\end{equation}
Using eq.~(\ref{eq:betam}) for the baryonic density profile we get,
\begin{equation}
   f_{esc} = 1-5.2c^3{4\pi \alpha_H^{(2)} \over 3 S_{tot}} f_g^2 \overline
   n_{vir}^2 R_{200}^3\;,
\end{equation}
and for the case $S_{tot} \propto M_b$,

\begin{equation}
   f_{esc} = 
\begin{cases}
1-0.55f_g{(1+z)^3 \over \epsilon } & \text{if $(1+z) < 1.22 (\epsilon/f_g)^{1/3}$},\\
0 & \text{if $(1+z) > 1.22 (\epsilon/f_g)^{1/3}$}.
\end{cases}
\label{eq:c1}
\end{equation}
When \fesc is not zero, the number of photons absorbed is such that
all the gas in the halo is kept ionized.  In Figure~\ref{fig:p1}, the
long-dashed lines show \fesc computed by using eq.~(\ref{eq:c1}) for
two values of $\epsilon =4, 40$. The solid lines are the fits to our
fiducial models for the same efficiencies and $M_{DM}=10^6, 10^7$
M$_\odot$. It is clear that, whatever stellar density profile we
assume, \fesc will not be less than the value given by
eq.~(\ref{eq:c1}) because the gas is not able to absorb more photons.
In our calculations we do not account for the effect of overlapping
H~II regions. Therefore, when the porosity of the ionized regions
approaches unity, we underestimate \fesc.  However, the collective
effects of multiple OB associations do not become important until
their H~II regions overlap (\ie, at high porosity parameter). As
shown by \cite{Dove:99}, this overlap is usually accompanied by the
production of a radiative shell, and the highest luminosity O stars
have faded by the time this shell breaks out. Thus, we do not believe
this to be a major effect.  A crude correction to the effects of H~II
region overlaps would be to use eq.~(\ref{eq:c1}) when $(1+z) < 1.22
(\epsilon/f_g)^{1/3}$.

In Figure~\ref{fig:p1} we show the effect on \fesc of different choices
of the stellar density distribution and radial cutoff. A general result is
that the stellar density distribution affects the dependence of \fesc
on the virialization redshift, while the dependence on the mass remain
basically the same, with \fesc decreasing as the mass increases.  If
the stellar density distribution follows the Schmidt Law ($\rho^*
\propto \rho_b$) and if we assume a stellar cutoff at $R=R_s$, we find
that \fesc is exactly the same as in our our fiducial models, where 
$\rho^* \propto R^{-2}$ and the same cutoff (solid line in
Figure~\ref{fig:p1} (left)).  If we set the stellar cutoff where the baryon
density is 1 cm$^{-3}$, we find that \fesc is approximately constant
at high-$z$ and increases steeply at low-$z$, approaching the values of
our fiducial models. At $z \sim 6$, when $R(n_b=1~{\rm cm}^{-3})=R_s$,
\fesc is the same as in our fiducial case and approximately equals
the constant value at high-$z$. This case is shown by the solid lines
in Figure~\ref{fig:p1} (left) for $\epsilon=4, 40$. If we set the cutoff at
$R=R(n_b=10~{\rm cm}^{-3})$, the solid line in Figure~\ref{fig:p1} (left),
the same reasoning holds with the only difference that the constant 
value of \fesc at high-$z$ is equal to the one at $z \sim 15$ in our 
fiducial case.
In Figure~\ref{fig:p1} (right) we show \fesc assuming a stellar density
distribution $\rho^* \propto \rho_b^2$ and a stellar cutoff at
$R=R(n_b=1~{\rm cm}^{-3})$. The effect of a more concentrated
distribution balances the increased cutoff radius at high-$z$,
producing the same \fesc as in our fiducial case.

Finally, we note that it is not unreasonable to have stars in the
outer part of the halo in high-$z$ objects. These halos are 
quite compact, and the crossing time is short compared to the typical
lifetime of an O or B star. These stars are therefore 
able to move substantial distances from their initial location. Using
eq.~(\ref{eq:r200}) and assuming a concentration parameter $c=10$,
we find $R_s = R_{\rm vir}/c \approx 70$ pc at 
$z_{vir}=10$ and $M_{DM}=10^7$ M$_\odot$. The
circular velocity of a halo, defined as $V_c=[G M(R) / R]^{1/2}$ is,
\begin{equation}
V_c=(3.5~{\rm km~s}^{-1})\left({\Omega_m h^2 \over
    0.15}\right)^{1/6}\left({M_{DM}
    \over 10^6~{\rm M}_{\odot}}\right)^{1/3}\left({1+z_{vir} 
    \over 10}\right)^{1/2}.
\end{equation}
If we assume that the typical dispersion velocity of the stars is
$V^*\sim V_c$, the crossing time is $t_{cross}=R_s/V_c \simeq (9.6
\times 10^6~{\rm yr}) [(1+z_{vir})/10]^{-1.5}$, independent of the
halo mass.  If the velocities of the gas or stars are bigger than
$V_c$, the DM halo will not be able to confine the baryonic matter and
the galaxy will be blown away, releasing metal-enriched gas and stars
into the IGM. If SNe explosions and stellar winds are active in this
object, the stellar and gas distributions can be strongly influenced
\citep{Ciardi:99}. In another possible scenario \citep{Parmentier:99},
the protogalaxy can survive to the explosion of several tens of SNe
because a significant part of the energy released by the SNe is lost
by radiative cooling. The associated blast waves trigger the expansion
of a supershell, sweeping all the material of the cloud and the
supershell is enriched by heavy elements. A second generation of star
is born in these compressed and enriched layers of gas. This second
generation cannot be suppressed by the Soft Ultraviolet Radiation
(SUVR), because the cooling is provided by metals, and will have a
very high \fesc due to the location in the outermost part of the halo
of the OB associations.

\subsection{Changing the Luminosity Function of OB Associations \label{ssec:lf}}

\def\cappb{%
  Escaping fraction as a function of the halo mass for different
  choices of the stellar luminosity function for $z_{vir}=6,
  \epsilon=40$. (left) The solid line shows \fesc for our fiducial model,
  the long-dashed line shows the effect of decreasing the upper cutoff
  to $S_2=10^{50}$ s$^{-1}$, and the dotted line shows the effect of
  decreasing the lower cutoff to $S_1=10^{47}$ s$^{-1}$. (right) The solid
  lines show \fesc for $\alpha=1$ and (from top to bottom)
  $S_2=10^{51}, 10^{50}, 10^{49}$ s$^{-1}$.  The dashed line show
  \fesc for $\alpha=3$.}  
\placefig{
\begin{figure}
\epsscale{1.1}
\plottwo{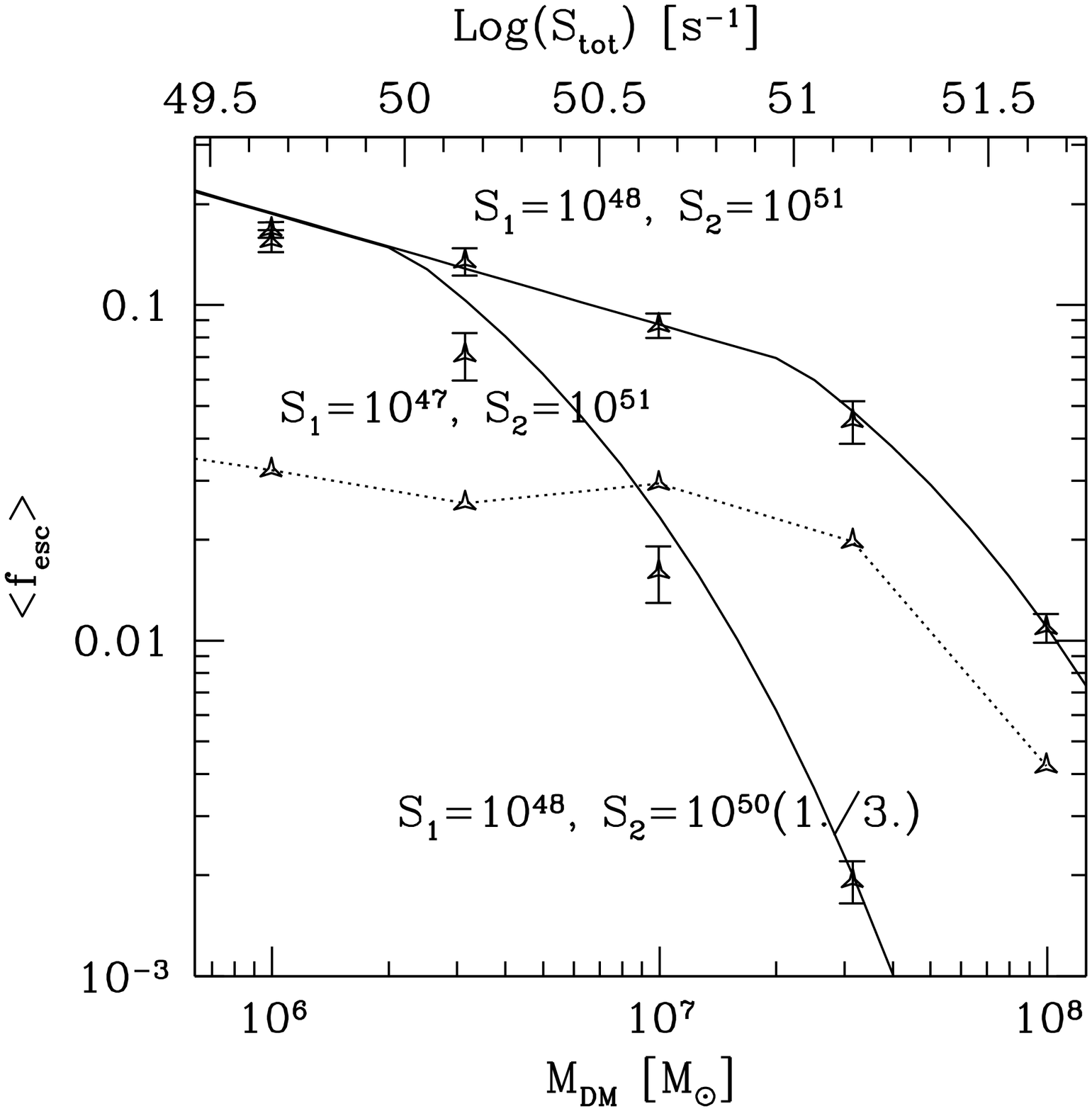}{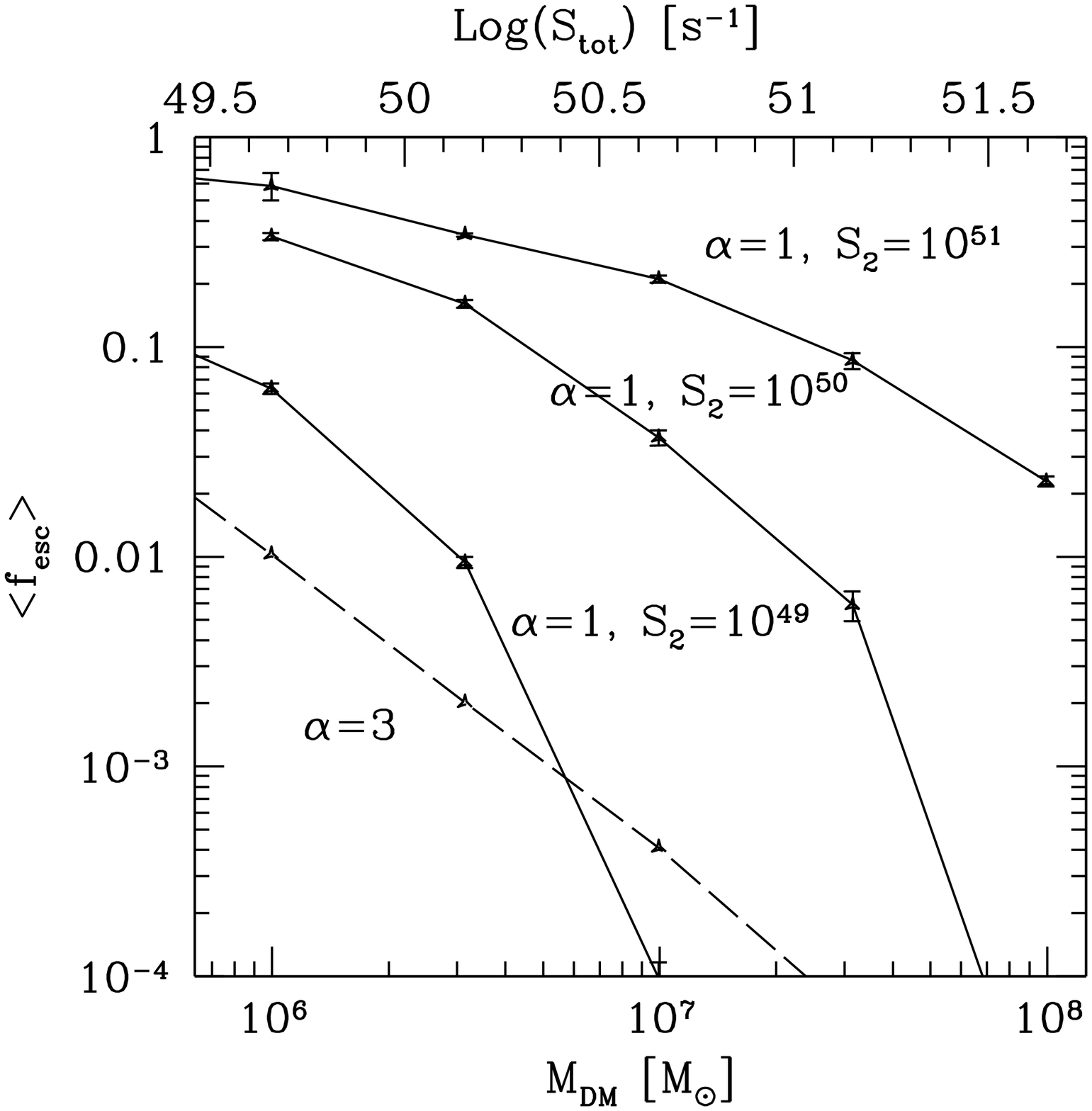}
\caption{\label{fig:p2}\cappb}
\end{figure}
}

In this section we study the dependence of \fesc on the adopted
luminosity function (LF) of the OB associations. We find that the
choice of the LF affects the dependence of \fesc on the mass of the
halo.  We adopt a power-law luminosity function as in
\S~\ref{sec:meth}, but we explore the effects on \fesc of different
choices of the slope $\alpha$ and the lower ($S_1$) and upper ($S_2$)
cutoffs of the LF.  In Figure~\ref{fig:p2} (left) we show \fesc for a
fixed slope $\alpha=2$ but changing the cutoffs. The solid line shows
\fesc when $S_1=10^{48}$ and $S_2=10^{51}$ s$^{-1}$ (fiducial model),
the long-dashed line when $S_1=10^{48}$ and $S_2=10^{50}$ s$^{-1}$, and
the dashed line when $S_1=10^{47}$ and $S_2=10^{51}$ s$^{-1}$. For
comparison, the solid line show \fesc for our fiducial case. \fesc is
on average lower if we chose smaller values of $S_1$ or $S_2$, and it
decreases more steeply with the mass when $S_2$ is decreased.

In Figure~\ref{fig:p2} (right) we show \fesc for different slopes of the LF:
$\alpha=3$ (dashed line) and $\alpha=1$ for $S_2=10^{51}, 10^{50},
10^{49}$ s$^{-1}$ (solid lines). \fesc on average decreases if the
distribution is steeper but the dependence on $M_{DM}$ becomes
shallower and vice versa. In the case $\alpha=3$ the relationship
between \fesc and $M_{DM}$ is a steeper power-law.  If $\alpha=3$, the total
luminosity is the sum of many ($N_{OB}=0.5(S_{tot}/S_1)$), low
luminosity associations. Instead if $\alpha=1$ the few 
($N_{OB}=(S_{tot}/S_2)\ln{S_2/S_1}$) very luminous associations
dominate the total luminosity.
  
The simulations show that OB associations with luminosities smaller
than a critical luminosity, $L_c(M_{DM})$, are too small to let any Lyc
radiation escape into the IGM unless they are far enough from the
center. The critical luminosity increases with the mass of the halo.
Therefore, in massive objects, only the high luminosity tail of the LF
contributes substantially to \fesc. This explains the decrease of \fesc
with the increasing mass of the halo, the dependence on the LF slope, and
lower and upper cutoffs. When the LF is steep, only small luminosities
OB associations close to the stellar density cutoff contribute to
\fesc. Therefore, the dependence on the mass is shallower.

\section{Simple Estimate of the Lyc Emissivity from Galaxies\label{sec:ps}}

In this section we estimate the integrated Lyc emissivity from
galaxies as a function of redshift. The main goal
is to understand the relative contribution on the emissivity from
small and big objects and estimate the SFE ($\epsilon$) required to
match the observed values.  In our simple estimate, the number of
photons per Mpc$^3$ per second emitted at redshift $z$ is given by,
\begin{equation}
n_{ph}=\int_{M_{min}}^\infty n(V_c,z)S_{tot}(M_{DM},\epsilon)
    \langle f_{esc} \rangle (M_{DM},\epsilon,z) d V_c,
\end{equation}
where $n(V_c,z)dV_c$ is the number density ($Mpc^{-3}$) of DM halos as
a function of circular velocity and redshift, $M_{min}(z)$ is the
minimum DM mass for which the gas is able to cool and collapse
according to \cite{Tegmark:96}.

To obtain $n(V_c,z)$ we use the Press-Schechter \citep{Press:74}
formalism following \cite{White:91}. We assume a $\Lambda$CDM power spectrum of
density fluctuations as in \cite{Efstathiou:92}, with $\Omega_m=0.3$,
$\Omega_\Lambda=0.7$ and $h=0.7$. The power spectrum normalization is
determined from COBE measurements of the Cosmic Microwave Background
(CMB) and yields $\sigma_8=1.02$.

We express the emissivity, $E$, in terms of the number of
emitted photons per hydrogen atom per Hubble time. According to
\cite{Miralda:00}, in these units, the emissivity at redshifts
$z=(2, 3, 4)$ should be $E=(14, 8.1, 4.3)$. These values of
the emissivity have been derived from the observed values of the mean
flux decrement at those redshifts according to \cite{Rauch:97}. The
formation of objects with $M_{DM}<M_{ion}$ (see \S~\ref{sec:res} for the
definition of $M_{ion}$), can be suppressed or delayed by
the SUVR produced by the first stars. The
SUVR can dissociate the molecular hydrogen, which is the primary coolant 
of gas at $T \leq 10^4$~K at very low metallicity. Stellar feedback 
affects small objects by suppressing star-formation but can also increases 
\fesc by blowing away the gas.  

The relative contribution of small objects 
to the integrated \fesc depends on the stellar luminosity function (see
\S~\ref{ssec:lf}).  It also depends 
on the number density of DM halos $n(V_c,z)$
which is a function of cosmological parameters and the power
spectrum of perturbations.
In Figure~\ref{fig:ps} we show the emissivity and the mean \fesc
(averaged over all halo masses) as a function of redshift (solid
lines). The long-dashed lines and dot-dashed lines are the
contributions to the total emissivity and mean \fesc from
objects with $M_{DM}<M_{ion}$ and $M_{DM}>M_{ion}$ respectively. 
Triangles are the values of $E$ from \cite{Miralda:00}. 
In Figure~\ref{fig:psfg} we show the mean \fesc as a function of
redshift for different values of $\epsilon$ and $f_g$.

\def\capps{%  
  (left) Emissivity $E$, in photons per hydrogen atom per Hubble time as
  a function of redshift. We show (see labels in the figure) the emissivity, and the emissivity if we set \fesc to unity.
  The dashed lines are for halo masses $M<M_{ion}$, dot-dashed lines
  for masses $M>M_{ion}$ and solid lines for all masses. The crosses
  are the expected values of the emissivity at $z=(2,3,4)$ according
  to \cite{Miralda:00}.
  (right) The mean \fesc as a function of redshift (solid line). The
  dashed and dot-dashed lines have the same meaning as explained
  above. For both these figures the efficiency is $\epsilon \simeq 16$ and $f_g=1$. 
   }  
\placefig{
\begin{figure}
\epsscale{1.1}
\plottwo{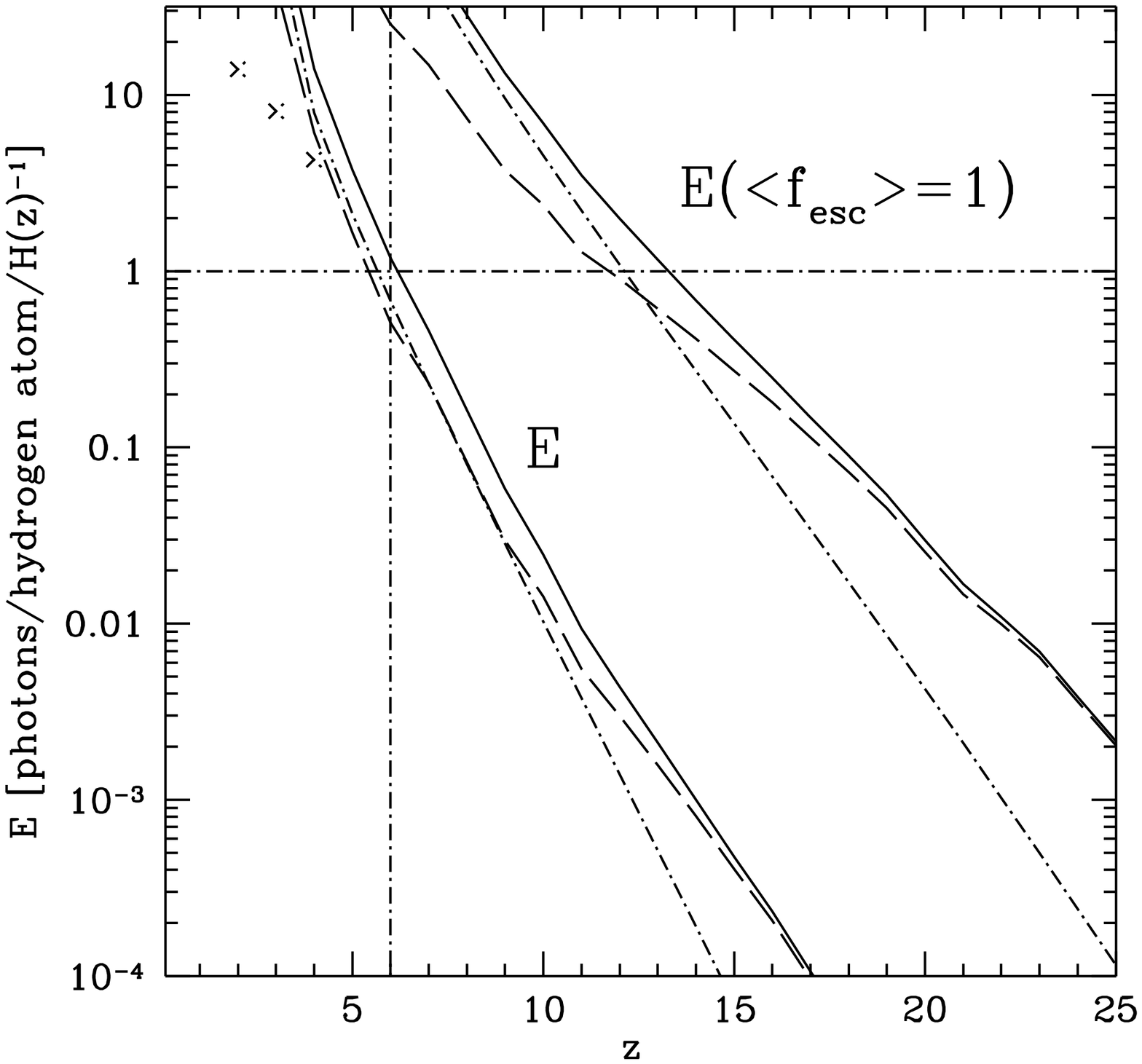}{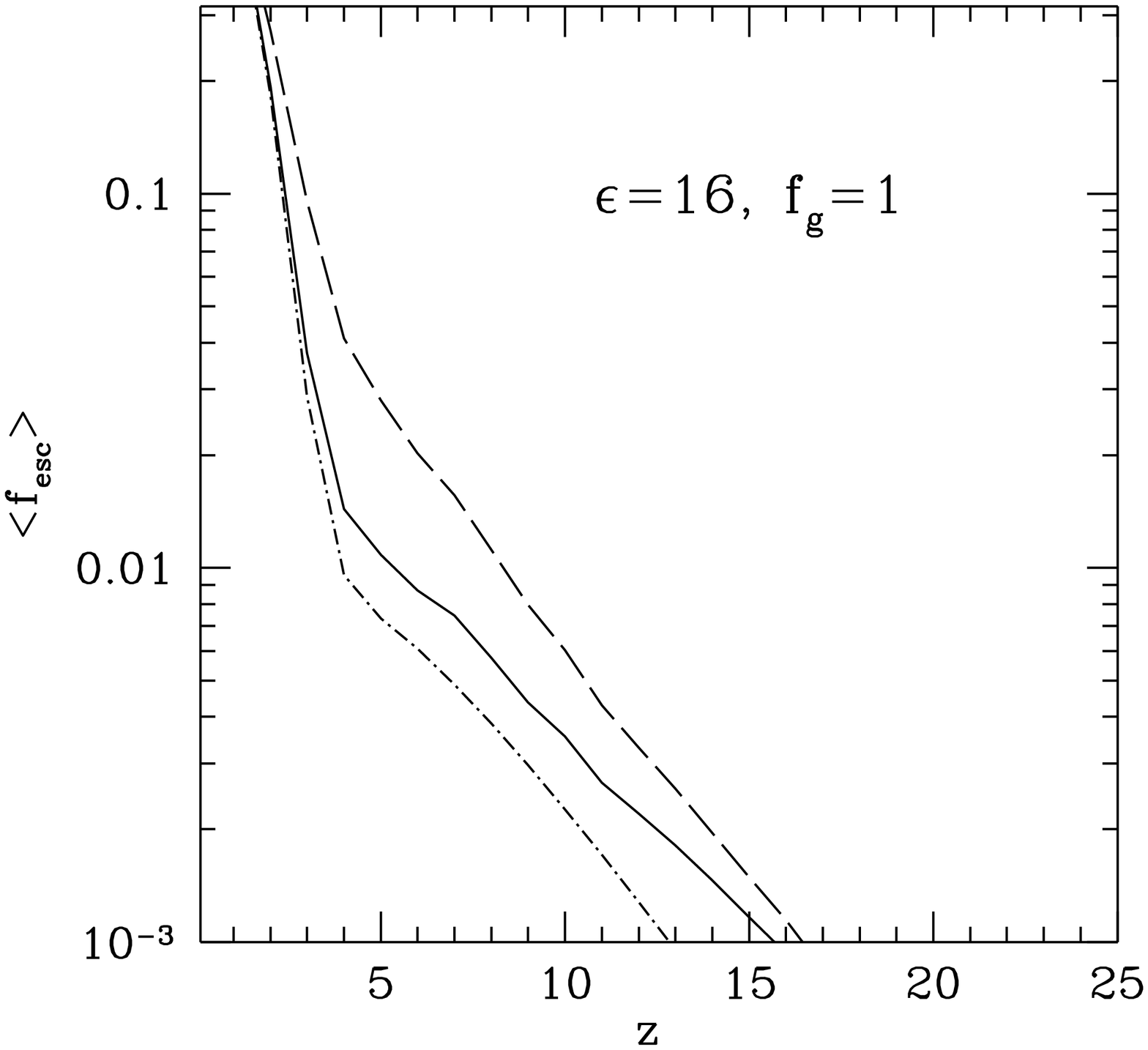}
\caption{\label{fig:ps}\capps}
\end{figure}
}
\def\cappsfg{%  
  The mean \fesc as a function of redshift for different values of
  $f_g$ and $\epsilon$ as indicated from the figure labels.}  
\placefig{
\begin{figure}
\epsscale{1.}
\plotone{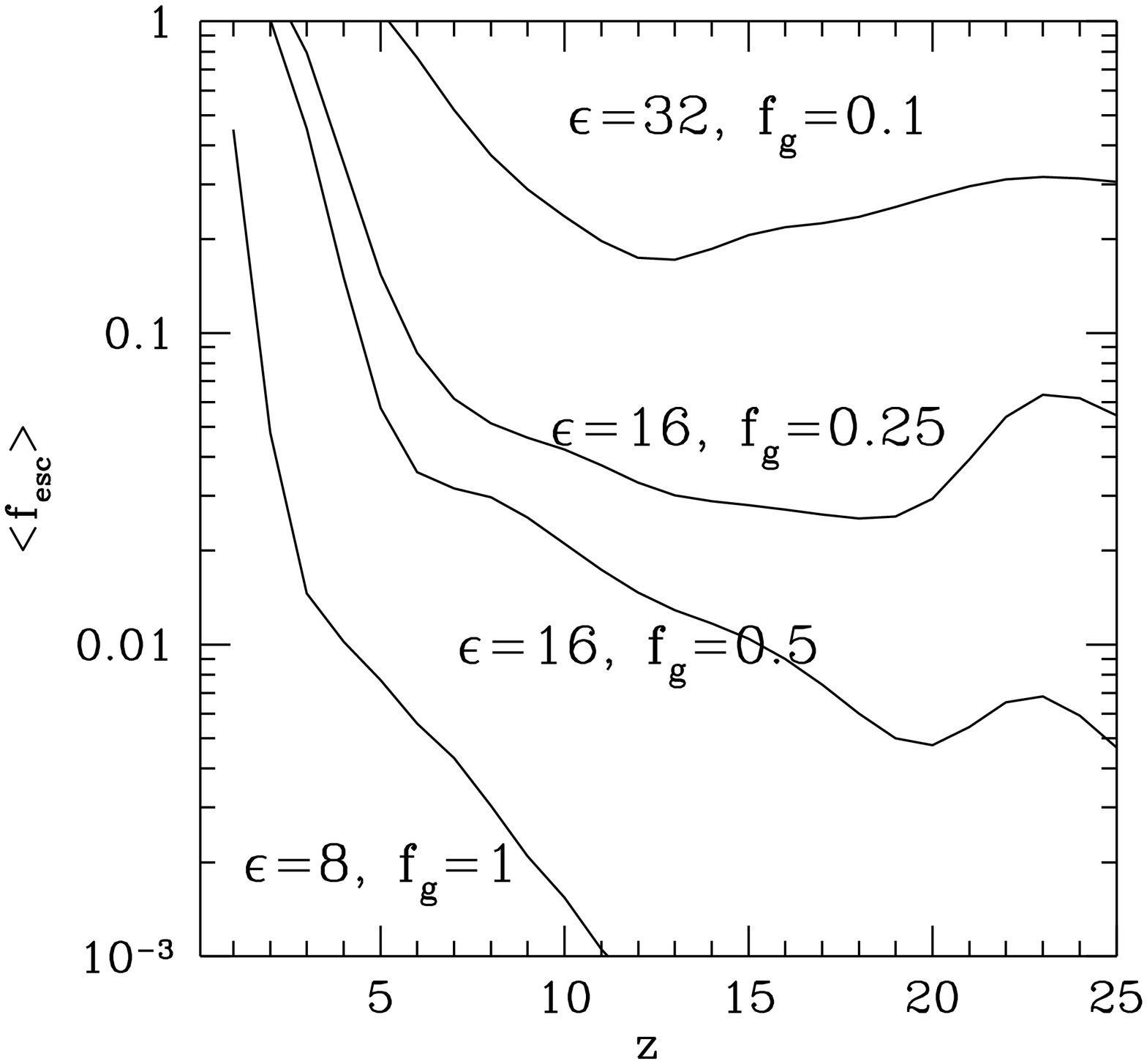}
\caption{\label{fig:psfg}\cappsfg}
\end{figure}
}

If \fesc does not depend on the halo mass (in Figure~\ref{fig:ps} we
show the case \fesc$=1$), the emissivity of the objects with
$M_{DM}>M_{ion}$ and $M_{DM}<M_{ion}$ would be equal at $z \sim 13$.
Instead, because of the dependence of \fesc on the halo mass, objects
with $M_{DM}<M_{ion}$ dominate the emissivity at $z\simgt 5$ if their
formation is not suppressed by feedback mechanisms.  A star formation
efficiency $\epsilon \sim 8$ is consistent with the observed
emissivity at $z \sim 4$.  However, if we assume that the reionization
occurs when $E \sim 1$, we find $z_{rei}\sim 5$. If the reionization
occurs at $z_{rei}\sim 7$ \citep{Gnedin:99, Gnedin:97} the star
formation efficiency needs to be $\epsilon \sim 16$. Within this
model, a constant (with redshift) star formation efficiency does not
reproduce the expected emissivity at $z=(2,3,4)$. In order to have a
good fit to the emissivity at $z=(2,3,4)$ and reionization at
$z_{rei}\sim 7$ the efficiency should be an increasing function of the
redshift. From observations of interacting galaxies,
\cite{Bushouse:99} have shown that the SFE increases with the amount
of molecular gas available in the galaxy.  Therefore, we expect
$\epsilon$ and $f_g$ to increase with redshift if, for a given
baryonic mass, the gas fraction increases with redshift. Finally we
remind the reader that our model assumes spherical galaxies and that
collisional ionization of the halo gas is negligible. Spherical or
irregular galaxies are probably numerous at redshift $z \simgt 2$
\citep{Dickinson:00} but at smaller redshift disk galaxies are
predominant. At low redshift the number of massive galaxies with a
collisionally ionized halo increases and the effect of dust absorption
also becomes more relevant. As a result the mean \fesc at $z \simlt 2$
should be calculated with other models (see \cite{Dove:99}).

\section{Summary and Conclusions\label{sec:sum}}

In this paper, we have investigated the ability of the first luminous
objects (Pop~III stars) in the universe to contribute to the ionizing
background that will reionize the universe between $z
\sim 10$ and 5. We consider a CDM cosmology, in which masses $M=10^6
- 10^7$~M$_{\odot}$ are able to collapse via $H_2$ cooling at
redshifts $z \simlt 30$. One of the most uncertain parameters is the
escape fraction, \fesc, of Lyc photons from the halo of the
galaxies. Knowing \fesc is crucial for determining the fraction of EUV
photons available to ionize the IGM. Unfortunately, the observations
of very high redshift galaxies is still beyond the capabilities of
ground-based and space-borne telescopes, and we can only begin to 
speculate about the physics of Pop~III objects \citep{Tumlinson:00}.

However, in many theoretical models of the IGM and reionization, an
estimate for the value of \fesc is needed. A typical value 
\fesc $\sim 10-20\%$ is usually adopted, based on both theoretical
studies \citep{Dove:94, Haiman:97} and on observations of local
starburst galaxies \citep{Leitherer:95, Hurwitz:97}. Both estimates
are valid for disk galaxies similar to the Milky Way at redshift 
$z \sim 0$. A recent study by \cite{Dove:99} that accounts for superbubble
dynamics finds \fesc$\simeq 0.03-0.06$.

At high redshift, the estimate of \fesc is even more uncertain because
of the lack of information on the SFE and IMF. For spheroidal galaxies,
the stellar density distribution can also affect \fesc. If the star
formation is triggered by protogalaxy merging, it is plausible that
many star-forming regions are located at the interface of the shock
waves produced by the collision. If the star formation is not a random
process, but is triggered by other star-forming regions, OB
associations will tend to form toward the outer boundary of the
galaxy, even if the first starburst happens in the center. In this
paper, we have investigated the effect on \fesc of several recipes
for the stellar density profile and luminosity function. The baryonic
density profile is calculated by assuming hydrostatic equilibrium in
the potential well of the DM halo. The gas effective temperature is
assumed to be equal to the virial temperature, and results from both
thermal and turbulent motions that support the gas.

Our key results are:

\noindent
(1) The escape fraction increases with the total number of Lyc photons
emitted per second, $S_{tot}$, and decreases with increasing mass and
redshift of the halo.

\noindent  
(2) The stellar ionization is dominated by small galactic objects.
For $S_{tot}=3.5 \times 10^{51}$~s$^{-1}$, we find \fesc $>10\%$ if
$M_{DM} \simlt 5 \times 10^7, 10^7, 2 \times 10^6, 10^6$~M$_{\odot}$
at $z_{vir}=3, 5, 10$, and 15 respectively. If we increase $S_{tot}$,
the DM masses listed above will increase approximatively by the same
amount. Therefore, at redshifts $z \simgt 10$, only galaxies with DM
halo masses $\simlt 10^7$~M$_\odot$ have \fesc $>10\%$ unless
the gas is collisionally ionized by shocks.

\noindent
(3) If we assume a SFE proportional to the baryonic content of the
galaxy, \fesc decreases exponentially with the redshift and as a $1/3$
power-law with the halo mass (if we set an upper cutoff for the
luminosity function of the OB associations \fesc decreases
exponentially for halo masses greater than a critical mass). The
dependency of \fesc on the redshift is related to the assumed stellar
density distribution and the dependency on the halo mass is related to
the OB association luminosity function.

\noindent
(4) We have found a simple analytical expression for \fesc as a
function of the normalized SFE, $\epsilon$, the redshift of virialization, $z_{vir}$,
and the DM halo mass, $M_{DM}$:
\begin{equation}
\log \langle f_{esc} \rangle = \log \left({min(Q,1)S_2 \over 10^2S_1}\right)^{-{1 \over
   3}}-0.41(1+\log f_g^{2 \over 3})\left({max(Q,1) \over \epsilon}\right)^{1 \over 3}(z_{vir}+1)\left({\Omega_m h^2 \over 0.15}\right)^{1/3},
\end{equation}
where $Q=S_{tot}/10S_2$. This formula is eq.~(\ref{eq:fit}) written in a
more compact form. 

\noindent
(5) A simple estimate of the emissivity as a function of redshift,
using the Press-Schechter formalism, shows that the emissivity is
dominated by small objects ($M_{DM} \sim 10^7$~M$_{\odot}$) up to
redshift 5 and that a SFE $\epsilon \sim 8$ is consistent with the observed
emissivity at $z=4$. A SFE $\epsilon \sim 15$ is needed to reionize the IGM at
$z_{ion} \sim 7$.

The question if these small objects with high SFE exist, has to be
addressed. Their formation can be suppressed by SNe explosions or by
Soft Ultraviolet radiation that prevent their cooling via $H_2$. On
the other hand, if they can survive to the SNe explosions from the
first generation of stars, a second generation can born on the
compressed and metal enriched gas layer produced by the blast waves.

Our study is a first attempt to understand the magnitude of \fesc from
a spheroidal galaxy as a function of redshift. Our results are a crude
approximation and can be improved by numerical simulations of galaxy
formation. In our treatment, we do not include the effect of dust
absorption, gas inhomogeneity, or gas dynamics. However, we believe
that adding further complications to the model is not justified until
observations tell us more about the nature of high-redshift
galaxies. This will allow us to build a more elaborate model based on
a solid observational ground.

\acknowledgements

This work was supported by the Theoretical Astrophysics program
at the University of Colorado (NSF grant AST96-17073 and NASA
grants NAG5-4312 and NAG5-7262).  We thank Mark Giroux for a
critical review of the manuscript and the anonymous referee for very
helpful suggestions.

%.......................................................................
\bibliographystyle{/users1/casa/ricotti/Latex/TeX/apj}
\bibliography{/users1/casa/ricotti/Latex/TeX/archive}

\begin{thebibliography}{}

\bibitem[\protect\citeauthoryear{{Abel} et~al.}{{Abel}
  et~al.}{1998}]{AbelAnninos:98}
{Abel}, T., {Anninos}, P., {Norman}, M.~L.,  \& {Zhang}, Y. 1998, \apj, 508,
  518

\bibitem[\protect\citeauthoryear{{Abel}, {Bryan}, \& {Norman}}{{Abel}
  et~al.}{1998}]{AbelBryan:98}
{Abel}, T., {Bryan}, G.,  \& {Norman}, M.~L. 1998, in Proceedings of MPA (ESO
  Conf. on "Evolution of Large Scale Structure", Garching) (astro-ph/9810215)

\bibitem[\protect\citeauthoryear{{Bennett} et~al.}{{Bennett}
  et~al.}{1994}]{Bennett:94}
{Bennett}, C.~L., et~al. 1994, \apj, 434, 587

\bibitem[\protect\citeauthoryear{{Binney}}{{Binney}}{1977}]{Binney:77}
{Binney}, J. 1977, \apj, 215, 483

\bibitem[\protect\citeauthoryear{{Boksenberg}}{{Boksenberg}}{1997}]{Boksenberg%
:97}
{Boksenberg}, A. 1997, in Structure and Evolution of the Intergalactic Medium
  from QSO Absorption Line System, 85

\bibitem[\protect\citeauthoryear{{Bushouse} et~al.}{{Bushouse}
  et~al.}{1999}]{Bushouse:99}
{Bushouse}, H.~A., {Lord}, S.~D., {Lamb}, S.~A., {Werner}, M.~W.,  \& {Lo},
  K.~Y. 1999, submitted (astro-ph/9911186)

\bibitem[\protect\citeauthoryear{{Ciardi} et~al.}{{Ciardi}
  et~al.}{1999}]{Ciardi:99}
{Ciardi}, B., {Ferrara}, A., {Governato}, F.,  \& {Jenkins}, A. 1999, submitted
  (astro-ph/9907189)

\bibitem[\protect\citeauthoryear{{Dickinson}}{{Dickinson}}{2000}]{Dickinson:00}
{Dickinson}, M. 2000, submitted (astro-ph/0004028)

\bibitem[\protect\citeauthoryear{{Donahue} \& {Shull}}{{Donahue} \&
  {Shull}}{1987}]{Donahue:87}
{Donahue}, M.,  \& {Shull}, J.~M. 1987, \apj, 323, L13

\bibitem[\protect\citeauthoryear{{Dove} \& {Shull}}{{Dove} \&
  {Shull}}{1994}]{Dove:94}
{Dove}, J.~B.,  \& {Shull}, J.~M. 1994, \apj, 430, 222

\bibitem[\protect\citeauthoryear{{Dove}, {Shull}, \& {Ferrara}}{{Dove}
  et~al.}{2000}]{Dove:99}
{Dove}, J.~B., {Shull}, J.~M.,  \& {Ferrara}, A. 2000, \apj, 531, in press
  (astro-ph/9903331)

\bibitem[\protect\citeauthoryear{{Efstathiou}, {Bond}, \& {White}}{{Efstathiou}
  et~al.}{1992}]{Efstathiou:92}
{Efstathiou}, G., {Bond}, J.~R.,  \& {White}, S. D.~M. 1992, \mnras, 258, 1P

\bibitem[\protect\citeauthoryear{{Fardal}, {Giroux}, \& {Shull}}{{Fardal}
  et~al.}{1998}]{Fardal:98}
{Fardal}, M.~A., {Giroux}, M.~L.,  \& {Shull}, J.~M. 1998, \aj, 115, 2206

\bibitem[\protect\citeauthoryear{{Gnedin}}{{Gnedin}}{1999}]{Gnedin:99}
{Gnedin}, N.~Y. 1999, submitted (astro-ph/9909383)

\bibitem[\protect\citeauthoryear{{Gnedin} \& {Ostriker}}{{Gnedin} \&
  {Ostriker}}{1997}]{Gnedin:97}
{Gnedin}, N.~Y.,  \& {Ostriker}, J.~P. 1997, \apj, 486, 581

\bibitem[\protect\citeauthoryear{{Haardt} \& {Madau}}{{Haardt} \&
  {Madau}}{1996}]{Haardt:96}
{Haardt}, F.,  \& {Madau}, P. 1996, \apj, 461, 20

\bibitem[\protect\citeauthoryear{{Haiman} \& {Loeb}}{{Haiman} \&
  {Loeb}}{1997}]{Haiman:97}
{Haiman}, Z.,  \& {Loeb}, A. 1997, \apj, 483, 21

\bibitem[\protect\citeauthoryear{{Hurwitz}, {Jelinsky}, \& {Dixon}}{{Hurwitz}
  et~al.}{1997}]{Hurwitz:97}
{Hurwitz}, M., {Jelinsky}, P.,  \& {Dixon}, W. V.~D. 1997, \apjl, 481, L31

\bibitem[\protect\citeauthoryear{{Kennicutt}, {Edgar}, \& {Hodge}}{{Kennicutt}
  et~al.}{1989}]{Kennicutt:89}
{Kennicutt}, R.~C., {Edgar}, B.~K.,  \& {Hodge}, P.~W. 1989, \apj, 337, 761

\bibitem[\protect\citeauthoryear{{Leitherer} et~al.}{{Leitherer}
  et~al.}{1995}]{Leitherer:95}
{Leitherer}, C., {Ferguson}, H.~C., {Heckman}, T.~M.,  \& {Lowenthal}, J.~D.
  1995, \apjl, 454, L19

\bibitem[\protect\citeauthoryear{{Lepp} \& {Shull}}{{Lepp} \&
  {Shull}}{1984}]{Lepp:84}
{Lepp}, S.,  \& {Shull}, J.~M. 1984, \apj, 280, 465

\bibitem[\protect\citeauthoryear{{Madau}}{{Madau}}{1998}]{Madau:98}
{Madau}, P. 1998, in Molecular Hydrogen in the Early Universe, ed.
  E.~{Corbelli}, D.~{Galli}, \& F.~{Palla}, Mem. S.A.It.

\bibitem[\protect\citeauthoryear{{Madau} \& {Shull}}{{Madau} \&
  {Shull}}{1996}]{Madau:96}
{Madau}, P.,  \& {Shull}, J.~M. 1996, \apj, 457, 551

\bibitem[\protect\citeauthoryear{{Makino}, {Sasaki}, \& {Suto}}{{Makino}
  et~al.}{1998}]{Makino:98}
{Makino}, N., {Sasaki}, S.,  \& {Suto}, Y. 1998, \apj, 497, 555

\bibitem[\protect\citeauthoryear{{Meiksin} \& {Madau}}{{Meiksin} \&
  {Madau}}{1993}]{Meiksin:93}
{Meiksin}, A.,  \& {Madau}, P. 1993, \apj, 412, 34

\bibitem[\protect\citeauthoryear{{Miralda-Escud\'e}, {Haehnelt}, \&
  {Rees}}{{Miralda-Escud\'e} et~al.}{2000}]{Miralda:00}
{Miralda-Escud\'e}, J., {Haehnelt}, M.,  \& {Rees}, M.~J. 2000, \apj, 530, 1

\bibitem[\protect\citeauthoryear{{Miralda-Escud\'e} \&
  {Ostriker}}{{Miralda-Escud\'e} \& {Ostriker}}{1990}]{Miralda:90}
{Miralda-Escud\'e}, J.,  \& {Ostriker}, J.~P. 1990, \apj, 350, 1

\bibitem[\protect\citeauthoryear{{Navarro}, {Frenk}, \& {White}}{{Navarro}
  et~al.}{1996}]{Navarro:96}
{Navarro}, J.~F., {Frenk}, C.~S.,  \& {White}, S. D.~M. 1996, \apj, 462, 563

\bibitem[\protect\citeauthoryear{{Navarro}, {Frenk}, \& {White}}{{Navarro}
  et~al.}{1997}]{Navarro:97}
{Navarro}, J.~F., {Frenk}, C.~S.,  \& {White}, S. D.~M. 1997, \apj, 490, 493

\bibitem[\protect\citeauthoryear{{Norman} \& {Ikeuchi}}{{Norman} \&
  {Ikeuchi}}{1989}]{Norman:89}
{Norman}, C.,  \& {Ikeuchi}, S. 1989, \apj, 345, 372

\bibitem[\protect\citeauthoryear{{Parmentier} et~al.}{{Parmentier}
  et~al.}{1999}]{Parmentier:99}
{Parmentier}, G., {Jehin}, E., {Magain}, P., {Neuforge}, C., {Noels}, A.,  \&
  {Thoul}, A.~A. 1999, in press (astro-ph/9911258)

\bibitem[\protect\citeauthoryear{{Peebles} \& {Dicke}}{{Peebles} \&
  {Dicke}}{1968}]{Peebles:68}
{Peebles}, P. J.~E.,  \& {Dicke}, R.~H. 1968, \apj, 154, 891

\bibitem[\protect\citeauthoryear{{Pei}}{{Pei}}{1995}]{Pei:95}
{Pei}, Y. 1995, \apj, 438, 623

\bibitem[\protect\citeauthoryear{{Press} \& {Schechter}}{{Press} \&
  {Schechter}}{1974}]{Press:74}
{Press}, W.~H.,  \& {Schechter}, P. 1974, \apj, 193, 437

\bibitem[\protect\citeauthoryear{{Rauch} et~al.}{{Rauch}
  et~al.}{1997}]{Rauch:97}
{Rauch}, M., et~al. 1997, \apj, 489, 7

\bibitem[\protect\citeauthoryear{{Rees} \& {Ostriker}}{{Rees} \&
  {Ostriker}}{1977}]{Rees:77}
{Rees}, M.~J.,  \& {Ostriker}, J.~P. 1977, \mnras, 179, 541

\bibitem[\protect\citeauthoryear{{Reimers} et~al.}{{Reimers}
  et~al.}{1997}]{Reimers:97}
{Reimers}, D., {K\"ohler}, S., {Wisotzki}, L., {Groote}, D.,
  {Rodriguez-Pascal}, A., P.,  \& {Wamsteker}, W. 1997, \aap, 327, 890

\bibitem[\protect\citeauthoryear{{Ricotti}, {Gnedin}, \& {Shull}}{{Ricotti}
  et~al.}{2000}]{Ricotti:00a}
{Ricotti}, M., {Gnedin}, N.~Y.,  \& {Shull}, J.~M. 2000, \apj, 534, 41

\bibitem[\protect\citeauthoryear{{Schaye} et~al.}{{Schaye}
  et~al.}{1999}]{Schaye:00}
{Schaye}, J., {Theuns}, T., {Rauch}, M., {Efstathiou}, G.,  \& {Sargent}, L.~W.
  1999, submitted (astro-ph/9912432)

\bibitem[\protect\citeauthoryear{{Shapiro} \& {Giroux}}{{Shapiro} \&
  {Giroux}}{1987}]{Shapiro:87}
{Shapiro}, P.~R.,  \& {Giroux}, M.~L. 1987, \apjl, 321, L107

\bibitem[\protect\citeauthoryear{{Silk}}{{Silk}}{1977}]{Silk:77}
{Silk}, J. 1977, \apj, 211, 638

\bibitem[\protect\citeauthoryear{{Songaila} \& {Cowie}}{{Songaila} \&
  {Cowie}}{1996}]{Songaila:96}
{Songaila}, A.,  \& {Cowie}, L.~L. 1996, \aj, 112, 335

\bibitem[\protect\citeauthoryear{{Tegmark} et~al.}{{Tegmark}
  et~al.}{1996}]{Tegmark:96}
{Tegmark}, M., {Silk}, J., {Rees}, M.~J., {Blanchard}, A., {Abel}, T.,  \&
  {Palla}, F. 1996, \apj, 474, 1

\bibitem[\protect\citeauthoryear{{Tumlinson} \& {Shull}}{{Tumlinson} \&
  {Shull}}{2000}]{Tumlinson:00}
{Tumlinson}, J.,  \& {Shull}, J.~M. 2000, \apjl, 528, L65

\bibitem[\protect\citeauthoryear{{White} \& {Frenk}}{{White} \&
  {Frenk}}{1991}]{White:91}
{White}, S. D.~M.,  \& {Frenk}, C.~S. 1991, \apj, 379, 52

\bibitem[\protect\citeauthoryear{{White} \& {Rees}}{{White} \&
  {Rees}}{1978}]{White:78}
{White}, S. D.~M.,  \& {Rees}, M.~J. 1978, \mnras, 183, 341

\bibitem[\protect\citeauthoryear{{Whitmore} et~al.}{{Whitmore}
  et~al.}{1999}]{Whitmore:99}
{Whitmore}, B.~C., {Zhang}, Q., {Leitherer}, C., {Fall}, S.~M., {Schweizer},
  F.,  \& {Miller}, B.~W. 1999, \aj, 118, 1551

\end{thebibliography}

\vskip 2truecm

\placefig{\end{document}}

\clearpage

\newcounter{figurecap}
\setcounter{figurecap}{0}

\begin{center}
\bf Figure Captions
\end{center}

\refstepcounter{figurecap}
Fig.\ \thefigurecap---\label{xfig}\capxfig

\refstepcounter{figurecap}
Fig.\ \thefigurecap---\label{density}\capden

\refstepcounter{figurecap}
Fig.\ \thefigurecap---\label{simulation}\capsim

\refstepcounter{figurecap}
Fig.\ \thefigurecap---\label{fig:mz}\capfmz

\refstepcounter{figurecap}
Fig.\ \thefigurecap---\label{fig:cut}\capcut

\refstepcounter{figurecap}
Fig.\ \thefigurecap---\label{f_z1}\capfz

\refstepcounter{figurecap}
Fig.\ \thefigurecap---\label{fig:5}\capf5

\refstepcounter{figurecap}
Fig.\ \thefigurecap---\label{fig:p1}\cappa

\refstepcounter{figurecap}
Fig.\ \thefigurecap---\label{fig:p2}\cappb

\refstepcounter{figurecap}
Fig.\ \thefigurecap---\label{fig:ps}\capps

\refstepcounter{figurecap}
Fig.\ \thefigurecap---\label{fig:psfg}\cappsfg

\clearpage
\tabone

\end{document}